\documentclass[pdf]{iucr}              % DO NOT DELETE THIS LINE
\RequirePackage{graphicx}
\usepackage{graphicx}
\usepackage{amsmath}

     \journalcode{J}              % Indicate the journal to which submitted
\begin{document}                  % DO NOT DELETE THIS LINE

     %------------------------------------------------------------------------
\title{A computational tool for symbolic derivation of the small angle scattering from complex composite structures}
\shorttitle{SEB: Scattering Equation Builder}

     % Authors' names and addresses. Use \cauthor for the main (contact) author.
     % Use \author for all other authors. Use \aff for authors' affiliations.
     % Use lower-case letters in square brackets to link authors to their
     % affiliations; if there is only one affiliation address, remove the [a].

\author[a]{Tobias W. J.}{Jarrett}{}
\cauthor[a]{Carsten}{Svaneborg}{zqex@sdu.dk}{}

\aff[a]{University of Southern Denmark, Campusvej 55, DK-5230 Odense M, \country{Denmark}}

     % Use \shortauthor to indicate an abbreviated author list for use in
     % running heads (you will need to uncomment it).

%\shortauthor{Soape, Author and Doe}

     % Use \vita if required to give biographical details (for authors of
     % invited review papers only). Uncomment it.

%\vita{Author's biography}

     % Keywords (required for Journal of Synchrotron Radiation only)
     % Use the \keyword macro for each word or phrase, e.g. 
     % \keyword{X-ray diffraction}\keyword{muscle}

%\keyword{keyword}

     % PDB and NDB reference codes for structures referenced in the article and
     % deposited with the Protein Data Bank and Nucleic Acids Database (Acta
     % Crystallographica Section D). Repeat for each separate structure e.g
     % \PDBref[dethiobiotin synthetase]{1byi} \NDBref[d(G$_4$CGC$_4$)]{ad0002}

%\PDBref[optional name]{refcode}
%\NDBref[optional name]{refcode}

\maketitle                        % DO NOT DELETE THIS LINE

\begin{synopsis}
Scattering Equation Builder (SEB) is a C++ library for symbolically deriving
form factors for composite structures built by linking sub-units to each other. 
\end{synopsis}

\begin{abstract}
Analysis of small angle scattering (SAS) data requires intensive modelling to infer and characterize the structures present in a sample. This iterative improvement of models is a time consuming process. Here we present the Scattering Equation Builder (SEB), a C++ library that derives exact analytic expressions for the form factor of complex composite structures. The user writes a small program that specifies how sub-units should be linked to form a composite structure and calls SEB to obtain an expression for the form factor. SEB supports e.g. Gaussian polymer chains and loops, thin rods and circles, solid spheres, spherical shells and cylinders, and many different options for how these can be linked together. In the present paper, we present the formalism behind SEB, and give simple case studies such as block-copolymers with different types of linkage and more complex examples such as a random walk model of $100$ linked sub-units, dendrimers, polymers and rods attached to surfaces of geometric objects, and finally the scattering from a linear chain of 5 stars, where each star is build by four diblock copolymers. These examples illustrate how SEB can be used to develop complex models and hence reduce the cost of analyzing SAS data.
\end{abstract}

     %-------------------------------------------------------------------------
     % The main body of the paper
     %-------------------------------------------------------------------------
     % Now enter the text of the document in multiple \section's, \subsection's
     % and \subsubsection's as required.

\section{Introduction}\label{sec:intro}

\onecolumn
\begin{figure}\label{fig:workflow}
    \includegraphics[width=0.8\columnwidth]{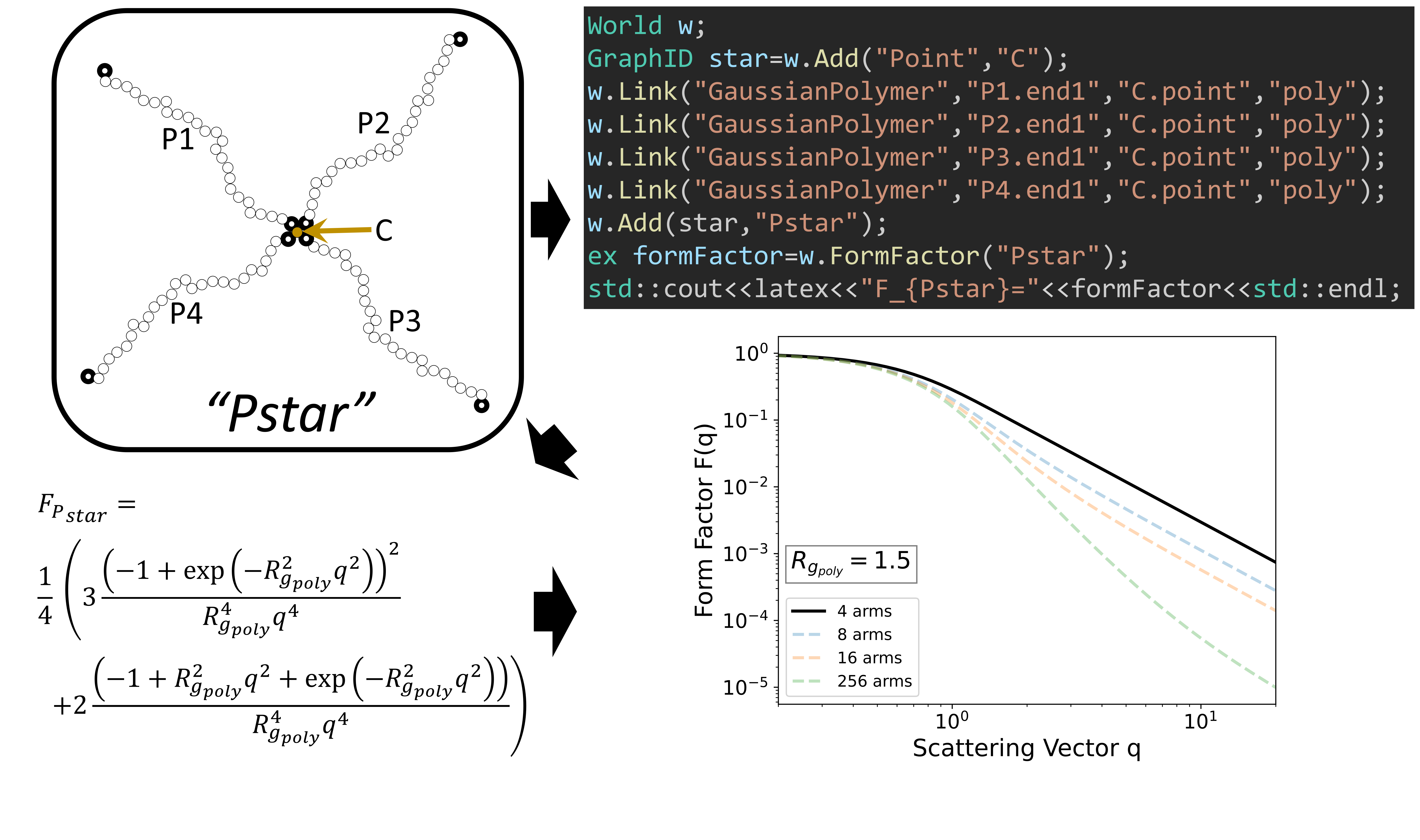}
    \caption{ SEB workflow: 1) defining a structure, 2) implementing the structure in SEB, 3) obtaining the analytic form factor equation, and 4) evaluating and plotting the form factor for given structural parameters.)
    }

\end{figure}
\twocolumn

\begin{figure}\label{fig:subunitoverview}
    \includegraphics[width=\columnwidth]{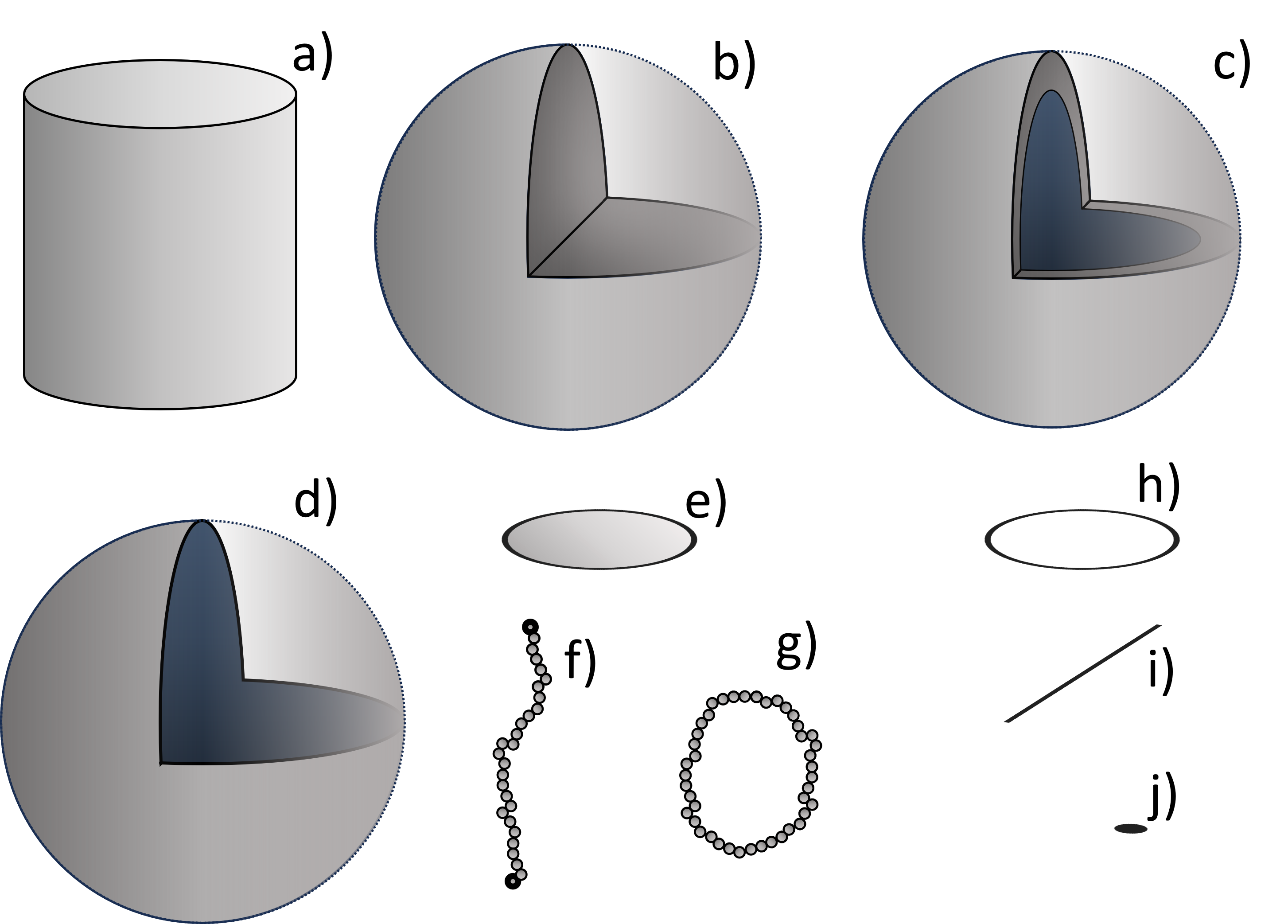}
    \caption{ Overview of supported sub-units
    a) a solid cylinder (SolidCylinder: center, hull, ends, surface),
    b) a solid sphere (SolidSphere: center, surface),
    c) a solid spherical shell (SolidSphericalShell: center, surfacei, surfaceo, surface),
    d) a thin spherical shell (ThinSphericalShell: center, surface),
    e) a thin disk (ThinDisk: center, surface, rim),
    f) a linear polymer (GaussianPolymer: end1, end2, middle, contour),
    g) a polymer loop (GaussianLoop: contour),
    h) a thin circle (ThinCircle: center, contour),
    i) a thin rod (ThinRod: end1, end2, middle, contour),
    j) an invisible point (Point: point). 
    For each sub-unit type, the parenthesis shows its SEB name and 
    reference points by which it can be linked to other sub-units.
    }
\end{figure}

Small angle scattering (SAS) is an ideal technique to characterize the size, shape and orientation of nano scale structures in a sample.\cite{guinier1955small,feigin1987structure} 
In order to infer the structures present in a sample, SAS scattering profiles are often analyzed by fitting models.\cite{JanAnalysis2} Thus SAS data analysis is an iterative process where models for structures are proposed, their corresponding scattering profiles are mathematically derived, and model scattering profiles are fitted to the experimental scattering profiles. If the fits are poor models have to be improved and the process starts over, until a good model has been developed. That is a model which provides an acceptable fit of the experimental data, and is thus the most likely candidate for the structures present in the sample.

SAS scattering spectra contain information about the nano scale structure, but not the detailed atomic scale structure, hence relatively simply geometric models models are often used when analyzing SAS data. Fortunately, the scattering from a large number of models has already been derived, see e.g. \cite{JanAnalysis2}. In the case where e.g. objects of similar shape are dispersed in a liquid, the problem of modelling the scattering from a sample can be split into 1) what are the shape of the objects, and 2) what are the spatial correlations of objects due to their mutual interactions.\cite{JanAnalysis2} The first problem is described by the form factor while the latter part is described by the structure factor, and in dilute samples the scattering is dominated by the form factor.

Here we present the Scattering Equation Builder (SEB), which is an C++ software library that analytically derives symbolic expressions for the form factor of composite structures built by linking an arbitrary number of sub-units together. A typical workflow is shown in Fig. \ref{fig:workflow}. Starting from a model of a structure (here a 4 arm star build of polymers), a short C++ program is written to define this structure within SEB. It typically takes SEB less than a minute to analytically derive a symbolic expression for the form factor of the structure. What is outside the scope of SEB is code required for representing and fitting experimental data or providing a graphical user interface. Already numerous excellent software tools has been developed with the specific aim of fitting models to scattering data for a non-exhaustive list see  e.g.
ATSAS\cite{konarev2006atsas,manalastas2021atsas},
CRYSOL\cite{svergun1995crysol},
CRYSON\cite{svergun1998protein},
FoXS/MultiFoXS\cite{schneidman2016foxs},
GENFIT\cite{spinozzi2014genfit},
GenApp\cite{perkins2016atomistic},
IRENA\cite{ilavsky2009irena},
SASfit\cite{bressler2015sasfit,kohlbrecher2022updates},
SASview\cite{doucet2017sasview}
Scatter\cite{forster2010scatter}, and
WillItFit\cite{pedersen2013willitfit}.

Our aim with SEB has been the ability to computationally efficiently derive form factor expressions for arbitrary complex branched structures. The expressions can be exported in a variety of formats allowing it to be imported into e.g. a C, C++, or Python programs, included LaTeX documents, or imported into Matlab or Matematica for further analysis. Finally, if the user specifies the length scales of the various sub-units, SEB can also evaluate the scattering equations to generate the corresponding scattering profile.

Fig. \ref{fig:subunitoverview} illustrates the sub-units that we have implemented in this initial release. The figure caption states which reference points we have implemented. These sub-units together with the large number of linkage options offered by the reference points define a large family of structures for which SEB can analytically derive scattering expressions.

SEB has been written in Object Oriented C++, which allows the expert user to expand SEB e.g. with additional sub-units and/or linkage options with relative ease. This choice also makes it possible to embed SEB within other software programs. SEB is Open Source and is freely available for download from GitHub at \cite{SEBgithub}. SEB depends on the GiNaC library\cite{bauer2002introduction} for internally for representing symbolic expressions, and GNU Scientific Library\cite{gough2009gnu} for evaluating certain special functions.

The paper is structured as follows, in Sect. \ref{sec:formalism} we briefly introduce the formalism and the logic behind SEB. SEBs design and implementation are presented in Sect. \ref{sec:seb}. Finally we present four advanced examples in Sect. \ref{sec:casestudies}. Sect. \ref{sec:summary} wraps up the article with a summary and outlook.

\section{Formalism}\label{sec:formalism}

We regard a composite structure as being created by linking sub-units together. For example, the structure of a semi-flexible polymer can be built by linking a sequence of rods end-to-end to form a linear chain of rods. The structure of a block-copolymer or a star-polymer can be built by linking two or more polymers together at one end. The structure of a di-block copolymer micelle can be built by linking polymers to the surface of a solid sphere representing the core. Here both the polymers and the sphere are sub-units. A bottle-brush polymer structure can be built by linking a number of short polymers to a random point along a long polymer chain.

Common for these example structures are that they are composites made of distinct sub-units linked in specific ways. Sub-units come in two varieties: simple geometric sub-units such as rods and spheres, and sub-units with internal conformational degrees of freedom such as polymers. In the latter case,  we need to perform conformational averages when predicting their scattering contributions.

For each type of sub-unit, we define specific reference points on a sub-unit where links can be made. For instance, a linear sub-unit such as a polymer or a rod has two distinct ends. These are points where we can link other sub-units. Each link represents the constraint that a reference point on one sub-unit is colocalized with a reference point on another sub-unit. A sphere can be linked to other sub-units at any random point on its surface. We could also imagine linking at any random point along the contour of the polymer or rod. This illustrates that reference points come in two varieties: specific geometric reference points such as the ends of a polymer or a rod, or distributed reference points such as random points on a geometric surface or along a polymer chain. When predicting scattering contributions, we also have to perform averages over distributed reference points. Even with e.g. a polymer sub-unit, we can link it together in many ways forming many structures e.g. block copolymers, star polymers, dendrimers or bottle brush structures or any combination of these.

To calculate the scattering from a composite structure we utilize the formalism of Svaneborg and Pedersen \cite{scatterformalism,scatterformalism2}. The formalism is based on three assumptions: 1) a structure does not contain sub-units that are linked into closed loops, 2) links are completely flexible, and 3) sub-units pairs are mutually non-interacting.
These three assumptions ensure that the internal conformation and orientation of all sub-units are statistically independent.
Interactions between different sub-units (3) would for instance create conformational correlations, for example in dense polymers the excluded volume interactions give rise to correlation hole effects in the scattering.\cite{schweizer1988integral}. When e.g. two rods are linked (2), the joint is flexible and can adopt any angle. If this was not the case, the links would create orientational correlations between the two rods. Finally if a structure contains loops (1), the closure constraint creates long range orientational and conformational correlations between all the sub-units involved in the loop.
When the internal conformation and orientation of all sub-units are statistically independent, the scattering from a composite structure can be factorized in terms of contributions from individual sub-units. No assumptions are made on the internal structure of sub-units and no additional assumptions or approximations are made. In this sense the formalism is exact.
SEB is an implementation of this formalism in C++. 
Below we introduce SEB and the formalism in more detail.

\subsection{Links}

\begin{figure}\label{fig:subunitlinks}
  \includegraphics[width = 0.99\columnwidth]{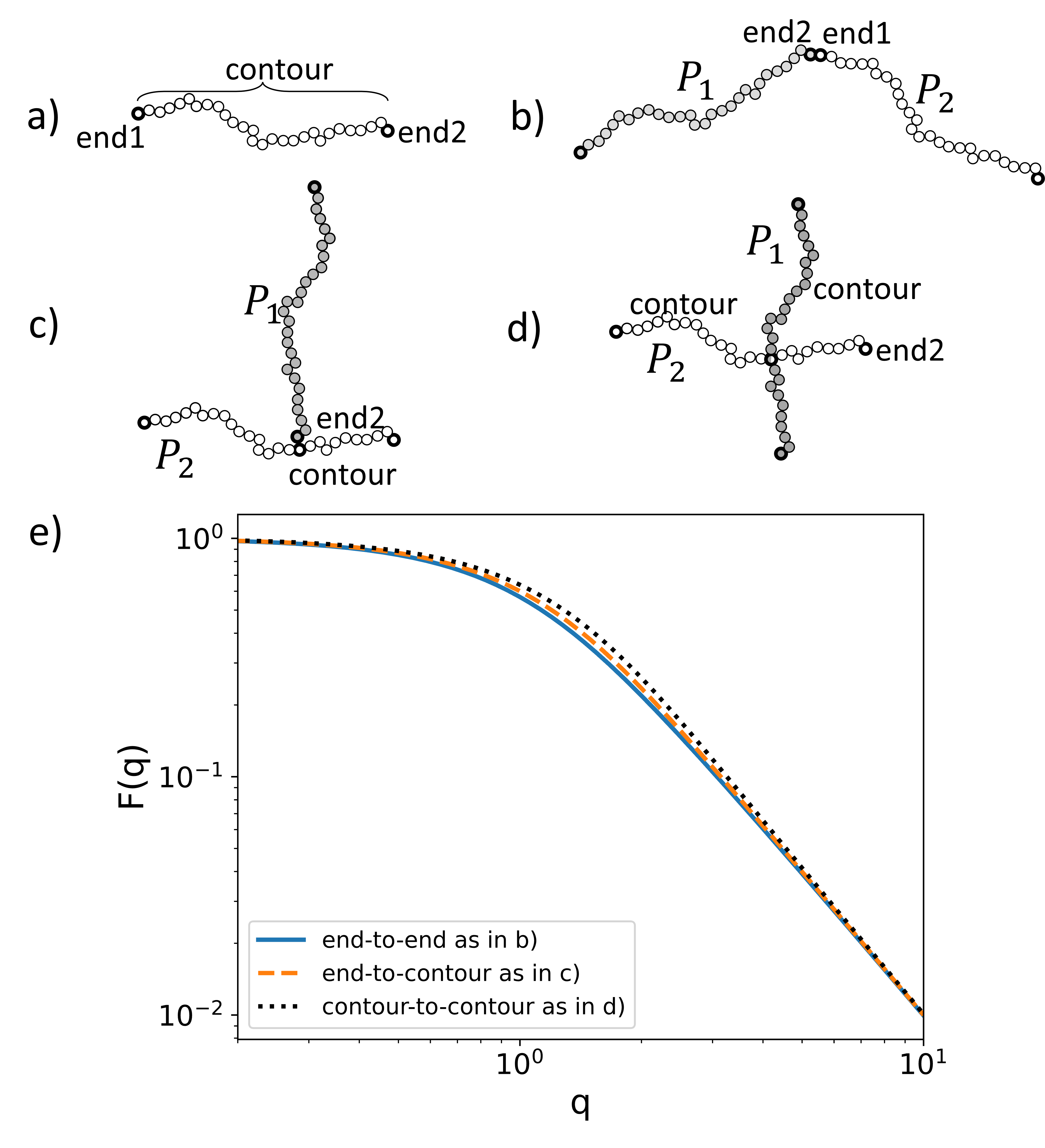}
\caption{

Illustration of a polymer sub-unit. (a) the three different reference points, (bcd) the three ways two polymers can be linked, (e) the scattering form factors for the different linkage options. 
}
\end{figure}

A sub-unit can have any number of specific and distributed reference points depending on its geometry. To keep track of them SEB
has hard coded labels for each reference point. For example, a polymer sub-unit has two specific reference points labeled "end1" and "end2", while it has one distributed reference point labeled "contour" (see Fig. \ref{fig:subunitlinks}a). 
Hence with just two polymers "P1" and "P2", we can create three different structures by linking "P1.end2" to "P2.end1" which produces a linear structure, "P1.end2" to "P2.contour" which produces a random 3-functional star structure, or "P1.contour" to "P2.contour" which produces a random 4-functional star structure.  Fig. \ref{fig:subunitlinks}bcd illustrates these structures.
When calculating scattering from structures with distributed reference points, we need to perform an average over random realizations of the link, hence we will obtain slightly different scattering profiles for these structures. Fig. \ref{fig:subunitlinks}e
shows the scattering form factor for these structures. In the Guinier regime observe that the radius of gyration is largest for the linear structure and smallest for the 4-functional star. At small $q$ values the structures produce the same scattering since they have the same scattering lengths, whereas for large $q$ values, we observe the power law scattering due to the internal random walk structure of the polymer, which is the same for all three structures.

\subsection{Sub-units}

A sub-unit is the building block of a structure. It is typically composed of many individual scatterers grouped together. We make
no assumptions about the internal structure of a sub-unit. Here and below we use capital latin letters to denote sub-units. The scattering contributions of the sub-unit is characterized by the following factors: The form factor is defined as
\begin{equation}\label{eq:subunitformfactor}
F_{I}(q) = \left(\sum_i \beta_i \right)^{-2}
\sum_{i,j} \beta_i \beta_j 
 \left\langle
\frac{\sin(qr_{ij})}{qr_{ij}}
\right\rangle,
\end{equation}
where  $r_{ij}=|{\bf R}_i-{\bf R}_j|$ is the spatial distance between the two scatterers, and $\beta_i$ denotes the excess scattering
length of the $i$'th scatterer. The form factor describes the interference contribution from all pairs of scatterers within the $I$'th sub-unit. Here and below we will use greek symbols to denote reference points.
For each reference point $\alpha$, the sub-unit has a corresponding form factor amplitude defined as
\begin{equation}\label{eq:subunitformfactoramplitude}
 A_{I\alpha}(\mathbf{q})=\left(\sum_j \beta_j \right)^{-1}
 \sum_j \beta_j 
 \left\langle
     \frac{\sin(qr_{j\alpha})}{qr_{j\alpha}}
\right\rangle,
\end{equation}
where $r_{j\alpha}=|{\bf R}_j-{\bf R}_\alpha|$ is the spatial distance between the $j$'th scatterer and the reference point.
The amplitude describes the phase difference introduced by the spatial distance between scatterers in a sub-unit and a reference point.
For each pair of reference points $\alpha,\omega$, the sub-unit has a corresponding phase factor defined as
\begin{equation}\label{eq:subunitphasefactor}
  \Psi_{I \alpha \omega}(q)=
  \left\langle  
\frac{\sin(qr_{\alpha\omega})}{qr_{\alpha\omega}}
\right\rangle,
\end{equation}
where $r_{\alpha\omega}=|{\bf R}_\alpha-{\bf R}_\omega|$ is the spatial distance between the two reference points.
The phase factor describes the phase difference between two specified reference points.

In these expressions, we have already performed the orientational average, however an additional average has potentially
to to be made over internal conformations and/or distributed reference points. For example, for a polymer described
by Gaussian chain statistics. For the "end1" form factor amplitude, one has to perform an average over the distribution
of distances between "end1" and any scatterer along the chain. For the "end1" to "end2" phase factor, one has to perform an average
of the polymer chain connecting the two ends. For the "contour" form factor amplitude of a polymer one has to perform a double
average over random position of the reference point along the chain and any scatterer along the chain. Finally for the
"contour" to "contour" phase factor, one has to average over two random positions of the reference points along the chain
as well as the Gaussian statistics of the polymer.

In the special case where distributed reference points (e.g. contour) and scatterers are characterized by the same distribution, such as a homogeneous distribution along the polymer, then the average expressions for the form factor amplitude and phase factor result in the same expression: the Debye expression for the form factor\cite{Debye}. We refer to Ref. \cite{scatterformalism,scatterformalism2} for the specific expressions.

\begin{figure}\label{fig:ellipses}
  \includegraphics[width = 0.99\columnwidth]{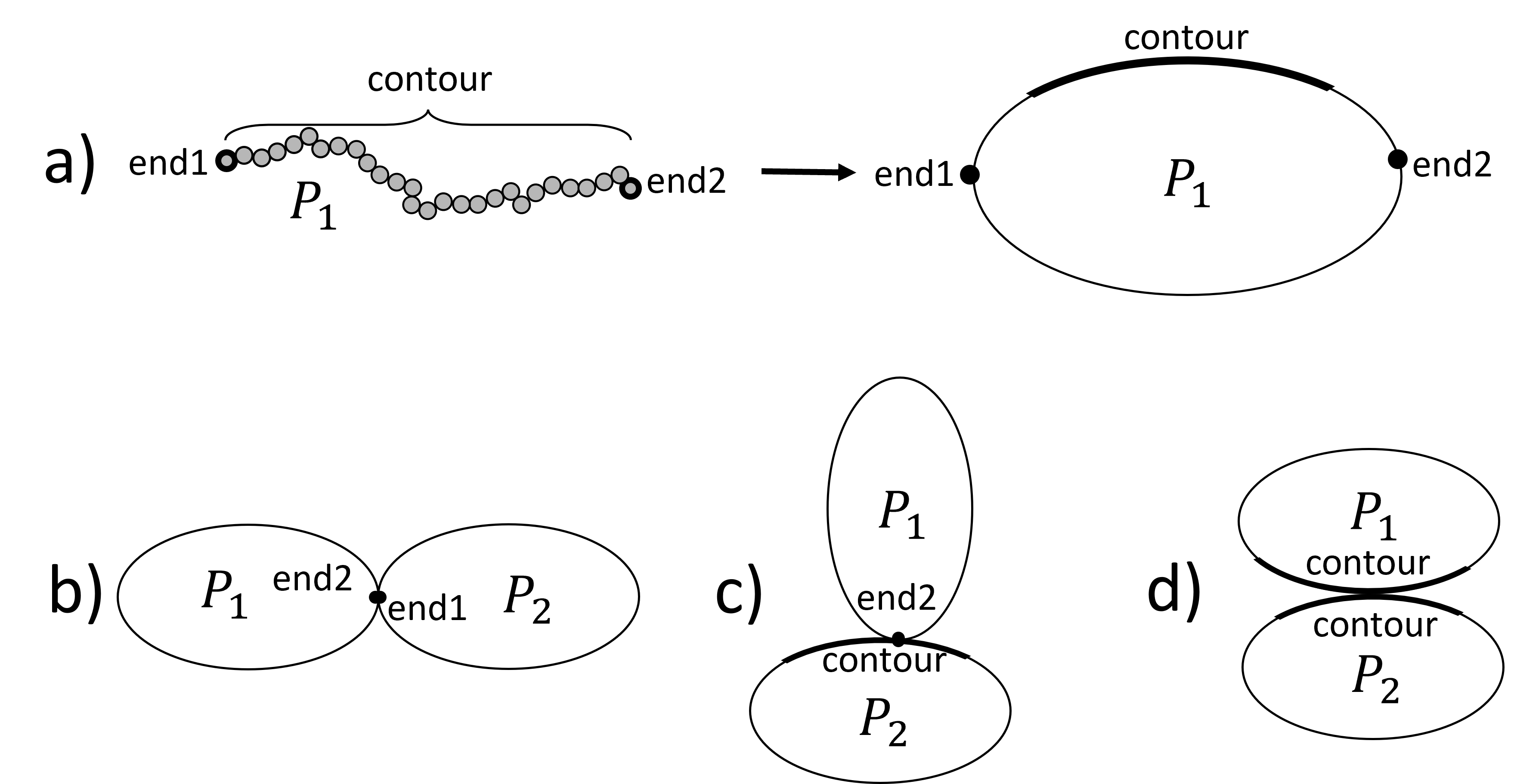}
  \caption{Illustration of (a) how a polymer and its reference points can be represented diagrammatically, and (bcd) how
  the different linkage options shown in Fig. \protect\ref{fig:subunitlinks} are represented.
}
\end{figure}

\begin{figure}\label{fig:possible-steps}
  \includegraphics[width = 0.80\columnwidth]{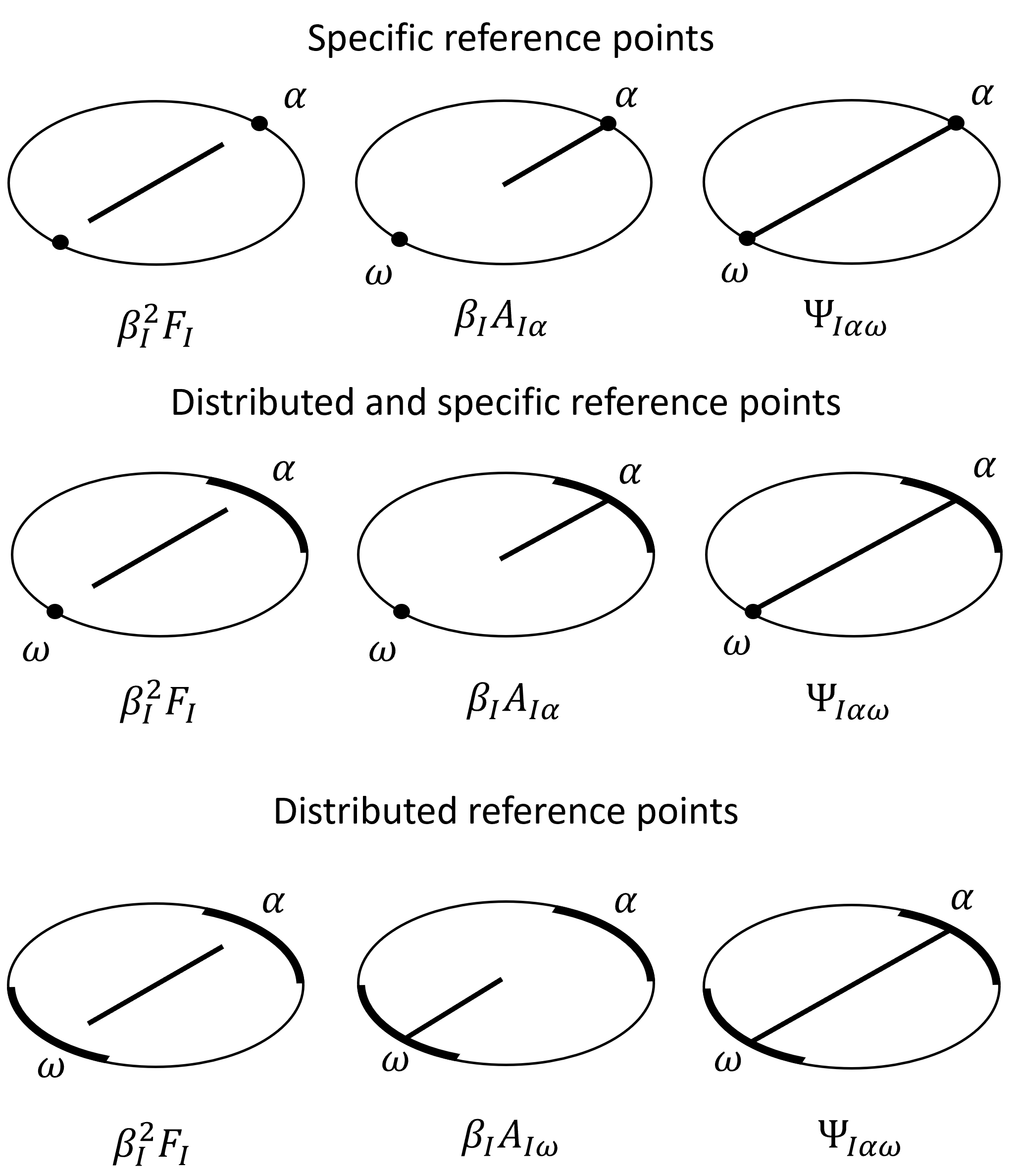}
  \caption{Library of all the possible diagrams and the corresponding factors to use when deriving scattering equations. }
\end{figure}

\begin{figure}\label{fig:composite}
  \includegraphics[width = 0.70\columnwidth]{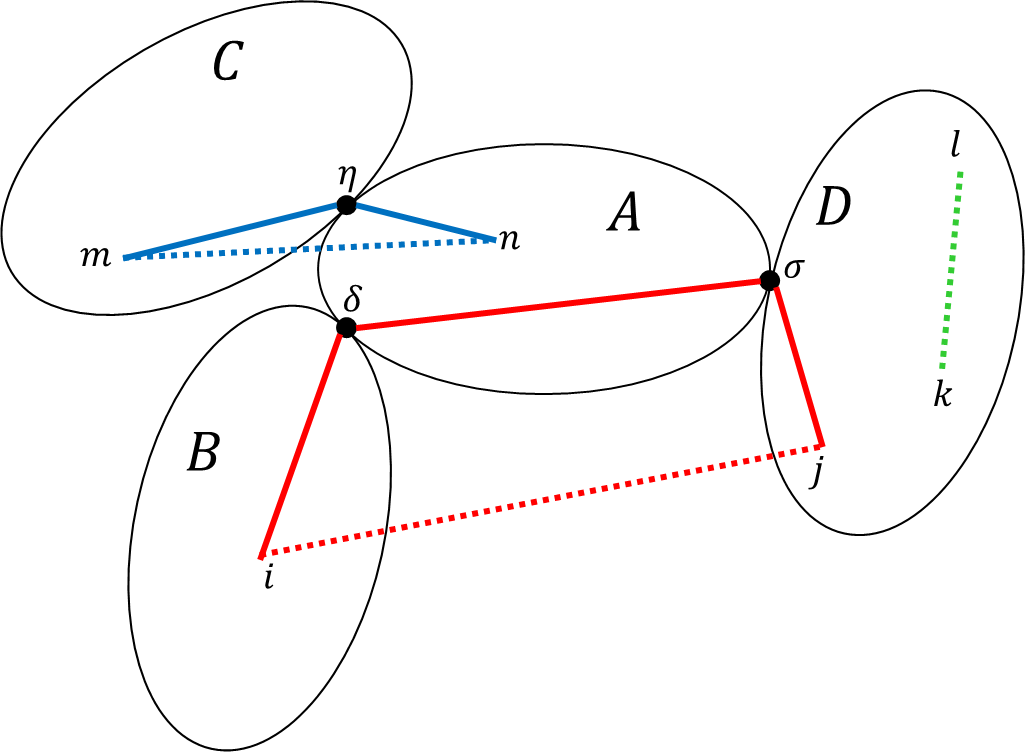}
  \caption{Example structure showing one sub-unit ($A$) with 3 pendant sub-units ($BCD$). The sub-unit are linked at three reference points ($\eta,\delta,$ and $\sigma$). Some scatterers within sub-units are illustrated as well (lower case letters). A few distances between scatterers are illustrated (colored dashed lines), together with their representations in terms of paths going through the structure (colored solid lines).
}
\end{figure}

\begin{figure}\label{fig:cornerstonestructure}
  \includegraphics[width = 0.70\columnwidth]{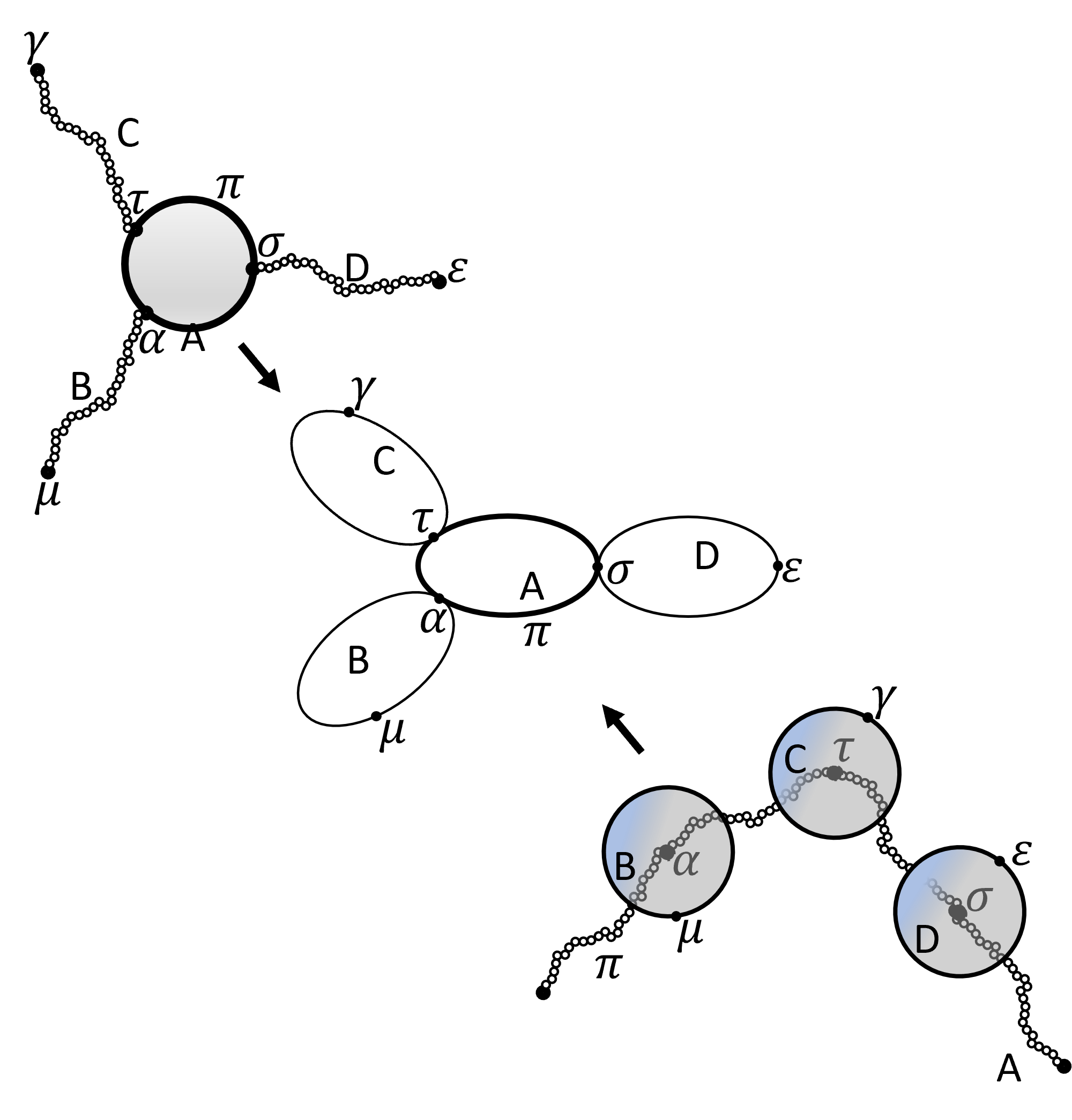}
  \caption{
Example structure build of three polymers linked to the surface of a sphere (top), three spheres linked by their center to the contour of a polymer (bottom), and the generic diagram with the same connectivity (center).
}
\end{figure}

\subsection{Diagrammatic interpretation}

A formal derivation of the general scattering expressions for a composite structure can be found in Refs. \cite{scatterformalism,scatterformalism2}. Before stating the general equations, we motivate the formalism with a diagrammatic derivation of the scattering from an example first.

To abstract from the concrete internal details of different sub-unit, we illustrate all sub-units by ellipses as shown in Fig. \ref{fig:ellipses}. Specific reference points are illustrated as dots on the circumference of the ellipse. Distributed reference points are illustrated as a thick line segment on the circumference of the ellipse to indicate that many points contribute. Fig. \ref{fig:ellipses}a shows a polymer and its diagrammatic representation. To illustrate links, the reference points on two sub-unit ellipses are shown as touching circumferences. The three linkage options shown in Fig \ref{fig:subunitlinks}bcd are illustrated in Fig. \ref{fig:ellipses}bcd. For simplicity, often we only show and label the reference points of interest when showing structures.

The total library of possible steps and the factors they contribute are shown in Fig. \ref{fig:possible-steps}. Diagrammatically form factors are derived from distances between pairs of scatterers within the same sub-unit, and hence they are illustrated as a line inside the ellipse. The form factor is also scaled by the square excess scattering length of the sub-unit. Form factor amplitudes are derived from distances between scatterers and a reference point, and they are illustrated by a line that starts inside the ellipse and ends on the circumference on the reference point. Form factor amplitudes are scaled by the excess scattering length of the sub-unit. Phase factors describe the phase introduced by the distance between two reference points, and hence are illustrated by a line between the two reference points. Since no scatterers are involved, phase factors do not depend on any excess scattering lengths.  Finally, when summing over all pairs of sub-unit we note that form factors are counted only once, however, all interference contributions are counted twice, since both the $I,J$ and $J,I$ paths contribute.

\subsection{Algorithm}

To calculate the form factor of a composite structure, SEB has to account for interference contributions between pairs of scatterers, while also keeping in mind that scatterers are grouped into linked sub-units. Fig. \ref{fig:composite} shows three illustrative cases, 1) the $l,k$ scatterers belong to the same sub-unit $D$, 2) the $n,m$ scatterers belong to directly linked sub-units $A,C$, and 3) scatterers $i,j$ belong to sub-units $BD$, that are indirectly connected via sub-unit $A$.

The first case of internal interference contributes between all scatterers within the same sub-unit is described by the form factor of the sub-unit $F_D$, here and below we suppress the dependency on $q$ for the sake of brevity. In the second case, the interference contribution between A and C depends on (average of) the vector $\Delta {\bf R} = {\bf R}_n-{\bf R}_m$, however stepping through the structure we note that $\Delta {\bf R}=({\bf R}_n-{\bf R}_\eta)+({\bf R}_\eta-{\bf R}_m)=\Delta {\bf R}_{n\eta} + \Delta {\bf R}_{\eta m}$. Where each parenthesis corresponds to an intra-sub-unit step. 
Since we have assumed that sub-units are uncorrelated, the spatial probability distribution of pair distances between scatterers in $P_{AC}({\bf R}_{n\eta},{\bf R}_{\eta m})$ can be written as a convolution of the two intra-sub-unit pair-distance distributions relative to the common reference point $P_A(\Delta {\bf R}_{n\eta})*P_C(\Delta {\bf R}_{\eta m}) $. In Fourier space, that convolution turns into the product of two sub-unit form factor amplitudes $A_{A\eta}A_{C\eta}$ both of which are evaluated relative to the common reference point $\eta$. This is the resulting interference contribution for case two.

Finally, the third case generalizes this logic. The interference contribution between scatterers $i,j$ depends on (average of) the vector $\Delta {\bf R}= {\bf R}_i-{\bf R}_j$. We note again that we can use reference points as stepping stones to write $\Delta {\bf R} = ({\bf R}_i - {\bf R}_\delta ) + ( {\bf R}_\delta+ {\bf R}_\sigma) + ( {\bf R}_\sigma-{\bf R}_j) = \Delta {\bf R}_{i\delta}+\Delta {\bf R}_{\sigma\delta}+\Delta {\bf R}_{\delta j}$. Each of the three parentheses describes an intra-sub-unit step. The distribution $P_{BAD}$ is a convolution of individual sub-unit contributions which factorizes into a product of three terms. However, since the middle step involves two reference points, hence the corresponding contribution is a phase factor. Thus the interference contribution becomes $A_{B\delta}\Psi_{D\delta\sigma}A_{D\sigma}$ for case three.

Hence the algorithm used by SEB for obtaining the scattering from a composite structure is to analyze all possible pairs of scatterers in the same or different sub-units. Hence the form factor is a double sum over all sub-units where we encounter three possible types of contributions: A form factor for scattering pairs belonging to the same sub-unit. Each pair of sub-units is either directly or indirectly connected. If they are directly connected, they contribute the product of their form factor amplitudes relative to the common reference point by which they are linked. If they are indirectly connected, we find the unique path through the structure connecting the two sub-units. This path uses reference points as stepping stones. The path is unique since the structure is assumed to be acyclic. The path contributes a form factor amplitude for the first and final sub-units relative to the first and final reference points in the path, respectively. Furthermore, each sub-unit along the path contributes a phase factor, which is to be calculated relative to the two reference points used to step across that sub-unit.

\subsection{Example}

\onecolumn
\begin{figure}\label{fig:cornerstoneexample}
  \includegraphics[width = 0.99\columnwidth]{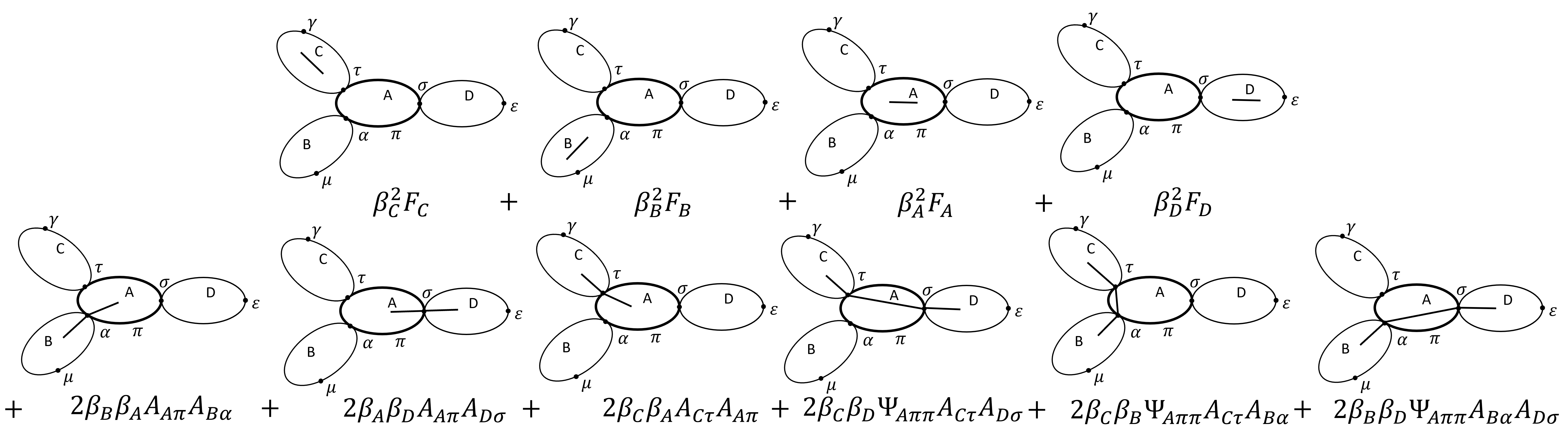}
  \caption{Diagrams of all the contributions to the form factor of an ABCD structure, where sub-units BCD are linked to sub-unit A.
  }
\end{figure}
\twocolumn

Fig. \ref{fig:cornerstonestructure} shows an example of a block-copolymer micelle modelled as three polymers
linked to the surface of a spherical core.\cite{PedersenGerstenberg} The figure also shows an example of three spheres linked by their center to a random position along contour of a polymer chain. This could be a beads-on-a-string model of a surfactant denatured protein.\cite{giehm2010sds} In the center of the figure, we show the diagrammatic representation where three sub-units are linked to a central sub-unit. We note that the generic diagram emphasizes the connectivity of the structure, and allows us to write down a generic equation for the form factor independent of the specific sub-units involved. In the figure, $\pi$ denotes the distributed reference point at which the other sub-units are linked.

For the simple example in Fig. \ref{fig:cornerstonestructure} we can enumerate all the possible scattering contributions from pairs of scatterers. This is shown in Fig. \ref{fig:cornerstoneexample} where the top, middle, and bottom rows, respectively, correspond to scattering pairs within the same sub-unit, within directly linked sub-units, and finally between indirectly connected sub-units, respectively. We note that all interference terms are counted twice since $IJ$ and $JI$ interferences contribute the same terms. Form factors only contribute twice. The reason is that while both $r_{ij}$ and $r_{ji}$ vectors between two scatterers $i,j$ contribute to the form factor, this is already accounted for by eq. (\ref{eq:subunitformfactor}). Summing all the scattering terms we get the (unnormalized) form factor of the structure. 

To finally derive the expression for a block copolymer micelle, we have to substitute the concrete polymer expressions for sub-units BCD, and the sphere expressions for sub-unit A. To finally derive the expression for the beads-on-a-string model, we instead substitute the concrete sphere expressions for sub-units BCD, and the polymer expressions for sub-unit A. These expressions can be found in Ref. \cite{scatterformalism2}.

When requesting the form factor of a structure in SEB, the user either obtains a generic structural equation like the one in Fig. \ref{fig:cornerstoneexample}, but the default is for SEB to perform all the sub-unit substitutions and returns a form factor equation for the specific choice of sub-units. For more complex structures, enumerating all the potential scattering contributions by hand is a very tedious and error prone process. SEB automates the process of identifying paths and tallying the corresponding factors.

\subsection{Generic equations}

\begin{figure}\label{fig:hierarchical}
  \center
  \includegraphics[width = 0.99\columnwidth]{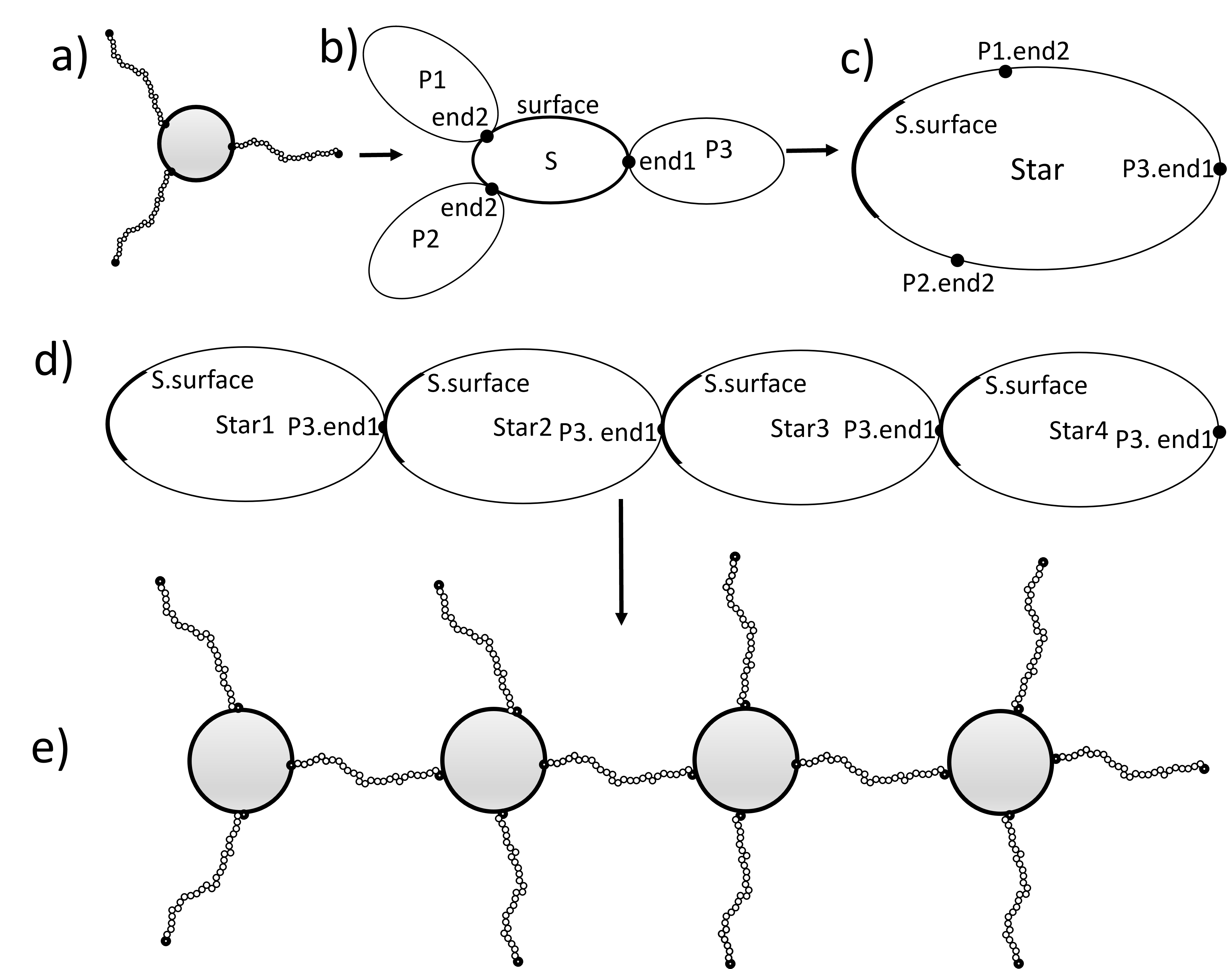}
  \caption{  
  Examples of hierarchical descriptions. A bottom-up description: a) a specific star structure made of sub-units linked to a core, b)
  the diagrammatic representation of sub-units in the star structure, and c) diagrammatic representation of a star sub-unit.
  A top-down description: d) four linked star sub-units, and e) the detailed structure when inserting the internal structure.
  }
\end{figure}

Just as a sub-unit has form factor amplitudes and phase factors, so does a composite structure that we have built out of sub-units.
Using the diagrammatic logic above, we can also draw the diagrams for form factor amplitudes of a structure relative to a reference point (not shown). In this case we have to sum over all sub-units in the structure. We find a path from the reference point to the sub-unit. The path contributes a product of phase factors for each sub-unit it traverses, and a form factor amplitude for the last sub-unit along the path relative to the last reference point. To calculate a phase factor of a structure relative to two reference points, we find the path through the structure connecting the reference points. The phase factor of the structure is the product of all the phase-factors of sub-units along that path. 

Generalizing the logic above, we can state the general expression for the form factor of a structure of sub-units. For each sub-unit pair $I,J$ we identify the first and final reference points $\alpha$ and $\omega$ and the path $P(\alpha,\omega)$ through the composite structure that connects them. Then the scattering interference contribution is the product of the form factor amplitudes of the first and final sub-units and of all the phase factors of sub-units along the path. The form factor of the composite structure is given by\cite{scatterformalism,scatterformalism2}

\begin{equation}\nonumber
  F_S(q) = \beta_{S}^{-2}\left[\sum_{I} \beta_{I}^{2} F_{I}(q) \right.
\end{equation}
  
\begin{equation}\label{eq:formfactorstructure}  
 \left.
  +\sum_{I \neq J} \beta_{I} \beta_{J}
  A_{I \alpha}(q) A_{J \omega}(q) \prod_{(K, \tau, \eta) \in \mathrm{P}(\alpha, \omega)} \Psi_{K \tau \eta}(q)\right],
\end{equation}

Having derived the form factor, it is straightforward to apply the same logic to state the equivalent form factor amplitude of a structure relative to any reference point it contains, as well as the phase factor of a structure relative to any reference point pair. These are given by\cite{scatterformalism,scatterformalism2}
\begin{equation}\label{eq:formfactoramplitudestructure}
  A_{S \alpha}(q) = \beta_{S}^{-1}\left[\sum_{I} \beta_{I} A_{I \omega}(q) \prod_{(K, \tau, \eta)
  \in \mathrm{P}(\alpha, \omega)} \Psi_{K \tau \eta}(q)\right],
\end{equation}
and
\begin{equation}\label{eq:phasefactorstructure}
  \Psi_{S \alpha \omega} =\prod_{(K, \tau, \eta) \in \mathrm{P}(\alpha, \omega)} \Psi_{K \tau \eta}(q).
\end{equation}

Usually the focus is on deriving form factors for different structures, and phase factors and form factor amplitudes are just intermediate expressions in the derivation. However, having all three scattering expressions for a structure allows us to use it as a sub-unit. In terms of mathematics, this corresponds to recursively inserting the left hand sides of eqs. (\ref{eq:formfactorstructure}-\ref{eq:phasefactorstructure}) into the right hand side of the equations.  
In terms of SEB, the code for generating scattering expressions makes recursive calls to itself until terminating at the sub-unit level. This hierarchical view of building structures using simpler sub-structures and sub-units as building blocks is a cornerstone of SEBs design. 

The logic is illustrated in Fig. \ref{fig:hierarchical}a-c, where the ABCD structure is wrapped into a single structure of type "star". In this case, we can think of e.g. "P1.end2" and "S.surface" as being the labels of reference points inside a star structure. In Fig. \ref{fig:hierarchical}d four instances of a star structure (named "star1"-"star4") are linked "P3.end1" to "S.surface". The resulting structure, a linear chain of stars, is shown in Fig. \ref{fig:hierarchical}e. With SEB, we would write code to link sub-units as in Fig \ref{fig:hierarchical}a, write a line to name the structure "star" thus realizing \ref{fig:hierarchical}c, and proceed to write code to build the structure \ref{fig:hierarchical}d using stars. Finally with a line of code we get the form factor of structure \ref{fig:hierarchical}e. Towards the end of the paper, we give an example where we build a diblock copolymer by joining two polymers. We then build a star by linking 4 diblock copolymers by one end, and proceed to build a chain where five stars are linked "tip" to "tip". This takes just 13 lines of code to do with SEB. Building hierarchical structures from more basic sub-structures vastly accelerates the time it takes to derive the scattering expressions.

Expressions for form factor amplitudes are also useful for modelling structure factor effects. If a structure has a reference point that could be regarded as the "center" of the structure, then SEB can also calculate the form factor amplitude relative to the center point $A_C$. In that case, an approximate model for the scattering including structure factor effects would be $I(q)=F(q)+A_C^2(q)(S_{CC}(q)-1)$, where $S_{CC}$ is a structure factor that describes the distribution of "center" to "center" distances between different structures.\cite{pedersen2001structure,pedersen2003small} This is analogous to the decoupling approximation\cite{KotlarchykChen} for polydisperse or anisotropic particles. The structure factor could e.g. be modelled as that of a hard-sphere liquid\cite{wertheim1963exact,thiele1963equation} or a hard-sphere liquid augmented with a Yukawa tail\cite{herrera1999static,cruz2008static}. Structure factor effects can also described using e.g. the RPA approximation\cite{Benoit} or using integral equation theory e.g. in the form of PRISM theory\cite{schweizer1987integral,curro1987theory,SchweizerCurro,DavidSchweizer,Yethiraj}. pyPRISM is a software package for numerically solving the PRISM equations\cite{martin2018pyprism}. We note that liquid state theories requires the form factor of a structure as an input, which can be derived with SEB.

\subsection{Estimating sizes}

While predicting scattering profiles is the main focus of SEB, we can also use analytic Guinier expansions of the scattering expressions to provide expressions for the size of composite structures. The size of a structure or a sub-unit can be gauged by three different measures: The radius of gyration $\langle R_g^2\rangle$ which describes the apparent mean-square distance between unique pairs of scatterers is obtained when expanding the form factor. The (apparent) mean-square distance between a given reference point and any scatterer $\langle R_{I\alpha}^2$ is obtained when expanding a form factor amplitude. Finally, the mean-square distance between a pair of reference points $\langle R_{I\alpha\omega}^2$ is obtained when expanding a phase factor.
We define the three Guinier expansions for a sub-unit $I$ as

\begin{equation}\label{eq:guinersexpansionF}
 F_I(q) \approx 1-\frac{q^2}{6}  \frac{ \left\langle \sum_{ij} \beta_i \beta_j r_{ij}^2 \right\rangle }{\left( \sum_{i} \beta_i \right)^2} +\cdots \equiv 1-\frac{2q^2 \left\langle R_{g,I}^2\right\rangle }{6}+\cdots
\end{equation}

\begin{equation}\label{eq:guinersexpansionA}
  A_{I \alpha}(q)
    \approx 1-\frac{q^2}{6}  \frac{ \left\langle \sum_{i} \beta_i r_{\alpha i}^2 \right\rangle }{\left( \sum_{i} \beta_i \right)} +\cdots 
  \equiv  1-\frac{q^2 \sigma_{I\alpha}\left\langle R_{I \alpha}^2\right\rangle }{6}+\cdots
\end{equation}

\begin{equation}\label{eq:guinersexpansionP}
  \Psi_{I \alpha \omega}(q) \approx 1-\frac{q^2 \left\langle  r_{\alpha \omega}^2 \right\rangle}{6}   +\cdots 
  \equiv  1-\frac{q^2 \sigma_{I\alpha\omega}\left\langle R_{I \alpha \omega}^2\right\rangle }{6}+\cdots
\end{equation}

Here the right hand side of the expressions defines the three size measures in terms of the expression in the middle. 
Based on the generic equations (\ref{eq:formfactorstructure}-\ref{eq:phasefactorstructure}),
we can derive three similar generic expressions for the size of any composite structure expressed in terms of the sizes of sub-units and paths through the structure. However, for simplicity we have directly implemented the Guinier expanded scattering terms for all sub-units in SEB, such that SEB explicitly calculates the Guinier expansion above (middle equations) and derives the size from the $q^2$ term in the expansion (right hand side).

Extra care has to be taken with regards to double counting of distances. The form factor includes the distance between any pair of scatterers twice since both $r_{ij}$ and $r_{ji}$ contribute to the form factor. We have made this double counting explicit by the prefactor of two in eq. \ref{eq:guinersexpansionF}. This has the effect of {\em defining} the radius of gyration from the {\em unique set of distances} between pairs of scatterers. For the form factor amplitude and phase factor, we occasionally have to account for a double counting. This done by introducing the double counting factors: $\sigma_{I\alpha}$ and $\sigma_{I\alpha\omega}$. 

In cases with specific reference points, pair distances between scatterers and reference points are unique by construction, and the double counting factor is unity. For instance, for the Guinier expansion of the form factor amplitude of a polymer relative to "end1", distances between "end1" and scatterers along the polymer are only summed once, hence $\sigma_{polymer,end1}=1$. Similarly, for the Guinier expansion of the phase factor between "end1" and "end2" of the polymer, the distance between the two ends of the polymer is summed only once, hence  $\sigma_{polymer,end1,end2}=1$.

In cases involving distributed reference points double counting can occur due to the additional average that has to be performed. For instance, Guinier expanding the form factor amplitude of a polymer relative to a "contour" reference point, we sum every distance between random points and scatterers twice, because both scatterers and reference points are uniformly distributed along the contour of the polymer. Hence $\sigma_{polymer,contour}=2$. Similarly, for the Guinier expansion of the phase factor between a pair of random "contour" points, we encountered every distance twice, hence $\sigma_{polymer,contour,contour}=2$ in this case as well. In fact, the set of distances between a random point on a polymer and a scatterer or between two random points on a polymer is exactly the same as the set of distances between pairs of scatterers, i.e. the mean-square distances from "contour" to scatterer and between two "contour" points is exactly the radius of gyration of the polymer. If we did not account for double counting in this case, we would have an inconsistency where e.g. the distance between randomly chosen points on a polymer would be twice the radius of gyration of the polymer. Note that SEB is not able to deduce whether double counting occurs in a given structure, hence SEB returns $\sigma_{I\alpha}\left\langle R_{I \alpha}^2\right\rangle$ and $\sigma_{I\alpha\omega}\left\langle R_{I \alpha \omega}^2\right\rangle$ to the user, and it is up to the user to divide the result by two in the rare cases where double counting has occurred.

\section{SEB}\label{sec:seb}

In the preceding section, we have illustrated the formalism. While its entirely possible to use the formalism to write down scattering expressions for complex structures by hand, this rapidly becomes tedious and error prone when many paths through a complex structure have to be enumerated, inserting the various expressions for sub-unit factors, and finally implementing the resulting expression in a SAS analysis software. 

The Scattering Equation Builder "SEB" is a Object Oriented C++ library that automates the process.
SEB calculates the form factor of a structure by identifying and traversing all the paths between unique sub-unit or sub-structure pairs. SEB can also calculate the form factor amplitude for a given reference point by exploring all the paths connecting that reference point to every other sub-units or sub-structures. Similarly, the phase factor between any two reference points is obtained by identifying the path between the reference points. In the case of hierarchical structures, the algorithm generates "horizontal" paths at a given structural level, and then evaluates scattering expressions by recursively exploring paths through sub-structures until the level of individual sub-units are reached. Internally, we have designed SEB to efficiently store a hierarchical graph representation of the structures and it uses efficient recursive algorithms to generate paths through the hypergraphs at a specified depth into the structure.

The SEB uses the GiNaC library\cite{bauer2002introduction} for representing symbolic expressions.
SEB depends on GNU Scientific Library\cite{gough2009gnu} for evaluating Sin integrals,
Bessel functions, and Dawson functions. SEB also includes code from J.-P. Morou\cite{JPMoreau}
for evaluating Struve functions.

The core functionality of SEB is to allow the user to write a short  program that
1) builds structures by linking specific uniquely named sub-units,
2) names a composite structure build by sub-units, such that it can be used as another sub-unit
3) builds hierarchical structures by linking simpler structures together,
4) to obtain analytic expressions characterizing the scattering and sizes of those structures
and/or 5) to save a file with a scattering profile for a chosen set of parameters.

From the user perspective, SEB exposes a very lean interface. Just four methods are
available for building structures. The user can choose to obtain generic structural scattering
expressions, expressions with all sub-unit scattering terms
inserted yielding an equation that depends explicitly on $q$ and a set of structural parameters.
The user can also obtain an intermediate representation, where scattering terms are inserted but
expressed with dimensionless variables, where all structural length scales are scaled by $q$.
Finally if the user defines the structural parameters and a vector of $q$ values, SEB can evaluate 
the scattering expressions to provide a vector of scattering intensities that can be saved to a file
for plotting.

Before going in detail with implementation and design choices, we start with two simple illustrative
examples: a diblock copolymer and a micelle / decorated polymer. 
These and more examples can be downloaded along with the SEB code from Ref. \cite{SEBgithub}.

\subsection{Diblock copolymer}

Creating a structure similar to the one seen in Fig. \ref{fig:subunitlinks}b involves a world to host the sub-units, and then creating two polymers and specifying how they are to be linked. The following complete C++ program does that
\begin{verbatim}
1:#include "SEB.hpp"
2:int main()
3:{
4: World w;
5: GraphID g=w.Add("GaussianPolymer", "A");
6: w.Link("GaussianPolymer", "B.end1", "A.end2");
7: w.Add(g, "DiBlockCopolymer");
8: cout << latex;
9: cout << w.FormFactor("DiBlockCopolymer");
10:}
\end{verbatim}
The first line includes the SEB header file, which declares what functions SEB provides. Lines 2-3, and 10 sets up the function main, which is executed when a program is run. Line 4 in the program creates an instance w of the World class. This instance provides all SEBs functionality to the user.

To create a structure in the world, we must first add and link the two polymers. In the fifth line, the user uses the w.Add() method to add a polymer to the world. "GaussianPolymer" refers to type of polymer described by a Gaussian chain statistics. With the second argument, the user assigns the unique name "A" to this sub-unit. The world returns a GraphID to the user in response to adding the sub-unit. The GraphID is a common ID shared by all sub-units linked together forming a graph.

In the third line, the user uses the w.Link() method to add and link a second GaussianPolymer sub-unit. With the second argument the user names this new sub-unit "B". With the second and third arguments the user defines that the new "B" should be linked by the "end1" reference point to "end2" on the already existing "A" sub-unit. To calculate the form factor and print it out, we must first wrap the graph formed by these two polymers in a structure. This is done in the fourth line with w.Add(), but this time it is called with a GraphID of the structure we want to name, and the string "DiBlockCopolymer". We note that all sub-unit and structure names are case-sensitive and  unique. Types of sub-units and their reference points names are hard coded in SEB (see Fig \ref{fig:subunitoverview}). Reference point names are also case-sensitive.

Having defined a structure in Lines 5-7, we now want to print out the equation for its form factor. The eight line specifies that we want the expression to be printed in the form of a LaTeX expression. With the command w.FormFactor( "DiBlockCopolymer") in the ninth line, the user requests the symbolic expression for the form factor. This is printed to the screen (cout $<$$<$). The form factor equation will be expressed in terms of the magnitude of the momentum transfer $q$, the structural parameters $Rg_A$, $Rg_B$, as well as the excess scattering lengths $\beta_A$, $\beta_B$. The names of the sub-units are used as subscripts in parameters used in the scattering expressions.

Here we chose LaTeX formatted output, but we could also have outputted the equation in formats compatible with C/C++, Python, or the native GiNaC format which is compatible with Mathematica / Matlab. GiNaC by default generates equations in expanded form and with a random unpredictable ordering of terms. This makes native latex formatted output lengthy. Most often we would export the scattering expression to a fit program, or to a symbolic mathematics program for simplification, or directly evaluate it to predict the scattering profile.

To change the diblock from "end2" to "end1" linking to random linking, such as Fig. \ref{fig:subunitlinks}c, we need to link "A.end2" to a randomly chosen point on "B.contour". Replacing line six with the following code snippet achieves that
\begin{verbatim}
6: w.Link("GaussianPolymer", "B.contour#r1", "A.end2");
\end{verbatim}

Here simultaneously with specifying the distributed reference point "contour" on the "B" sub-unit, we also label that (now specific) reference point with the arbitrary string "r1". If we instead want to create the structure of Fig. \ref{fig:subunitlinks}d, we need to link one random reference point on "B.contour\#r2" to a random reference point on "A.contour\#r3". Replacing line three with the following code snippet achieves that

\begin{verbatim}
6: w.Link("GaussianPolymer", "B.contour#r2", "A.contour#r3");
\end{verbatim}

The scattering profile corresponding to fig. \ref{fig:subunitlinks}bcd is shown in Fig. \ref{fig:subunitlinks}e. The difference is not large, but illustrates the point that even with the same sub-units different linkage options affect the scattering profile. The reference point name "contour" is hard coded in SEB, but the user is free to choose the labels (here "r1","r2","r3"). Having a unique name for each reference point allows us to add more sub-units to the same random point. Having both options for linking allows the user to develop well defined arbitrarily complex branched structures of end-to-end linked polymers, or bottle brush structures where many side chains are randomly attached to a main polymer.

As default SEB express scattering expressions in terms of an explicit $q$ value, a set of structural parameters and excess scattering lengths. The default option is also to output normalized scattering expressions such that they converge to unity in the limit of small $q$ values.  Replacing w.FormFactor( "DiBlockCopolymer") by w.FormFactorAmplitude( "DiBlockCopolymer:A.end1") would generate the form factor amplitude expression for the whole DiBlockCopolymer, but expressed relative to the specified reference point. With w.PhaseFactor( "DiBlockCopolymer:A.end1", "DiBlockCoPolymer:B.end2") SEB would instead generate the phase factor of the DiBlockCopolymer relative to the two specified reference points. With w.FormFactorGeneric( "DiBlockCopolymer") we would get the generic form factor of a structure of two connected sub-units without the specific scattering expressions inserted, this is often useful for debugging. Finally, with w.RadiusOfGyration2( "DiBlockCopolymer") SEB would generate the expression for the radius of gyration.

\subsection{Diblock copolymer micelle}

SEB is not limited to using one type of sub-unit type, but we can use and link all types of sub-units to each other. 
We can, for instance, model a diblock copolymer micelle as a number of polymer chains attached to the surface of a spherical core.\cite{PedersenGerstenberg} Here we limit the number of polymers to three for the sake of simplicity. To generate the micelle shown in Fig. \ref{fig:cornerstonestructure} (top), we need to create a solid sphere ("A") and add three polymers ("B", "C", and "D") to its surface, the following code snippet does that

\begin{verbatim}
1: World w;
2: GraphID g=w.Add("SolidSphere","A","s");
3: w.Link("GaussianPolymer","B.end1","A.surface#p1","p");
4: w.Link("GaussianPolymer","C.end1","A.surface#p2","p");
5: w.Link("GaussianPolymer","D.end1","A.surface#p3","p");
6: w.Add(g, "Micelle");
\end{verbatim}

A polymer sub-unit (type GaussianPolymer) has "end1", "end2", and "contour" as reference points, a solid sphere sub-unit (type SolidSphere) has "center" and "surface" as reference points. Just as we need to add labels for random points on the contour of the polymer above, we also add labels for the random points on the surface of the sphere. If we used the same label in all three Link commands, the three polymers would be linked to the same random point. This would influence the scattering interference between the polymers and is not the structure we are aiming to create. 

We also introduce tags in the example, which are an optional parameter of w.Add() / w.Link(). We tag all polymers as "p", and the spherical core as  "s". The result is that the scattering expressions are not stated in terms of the unique names A,B,C, and D, but are stated using the radius of gyration of the polymers $Rg_p$, and radius of the sphere $R_s$ as well as the two excess scattering lengths $\beta_p$ and $\beta_s$. If a tag is not specified, then the unique name is used in its place as in the diblock example above. By specifying tags, we can mark a set of sub-units as being identical in terms of their scattering properties and structural parameters.

\subsection{Decorated polymer}

A model of a surfactant denatured protein could be a long polymer with some spherical surfactant micelles along its contour. To generate a polymer decorated by three spheres  as in Fig. \ref{fig:cornerstonestructure} (bottom), we would use the following code snippet
\begin{verbatim}
1: World w;
2: GraphID g=w.Add("GaussianPolymer","A", "p");
3: w.Link("SolidSphere","B.center","A.contour#p1","s");
4: w.Link("SolidSphere","C.center","A.contour#p2","s");
5: w.Link("SolidSphere","D.center","A.contour#p3","s");
6: w.Add(g, "DecoratedPolymer");
\end{verbatim}

We note that this is nearly identical to the micelle code above, since we link three sub-units to a single sub-unit in both cases. The only difference being that instead of linking three polymers to a sphere, we link three spheres to one polymer. The three spheres "B", "C", and "D" are tagged with "s". Such that the scattering expression depends on the same parameters as described above.

\section{Advanced examples}\label{sec:casestudies}

Having discussed the basics of how to add and link sub-units, create structures, and output GiNaC expressions, here we show
how to implement some of the more advanced examples. In particular, we show a complete example of to write a program that
generates the scattering from $100$ identical linked sub-units for a variety of sub-units and linkage options, how to
generate a dendritic structure of linked sub-units, an example of polymers and rods linked to the surfaces of different
solid geometric objects, and finally how to implement a chain of 5 linked di-block copolymer stars using hierarchically defined building blocks.

\onecolumn
\begin{figure}\label{fig:chain100}
    \includegraphics[width = 0.80\columnwidth]{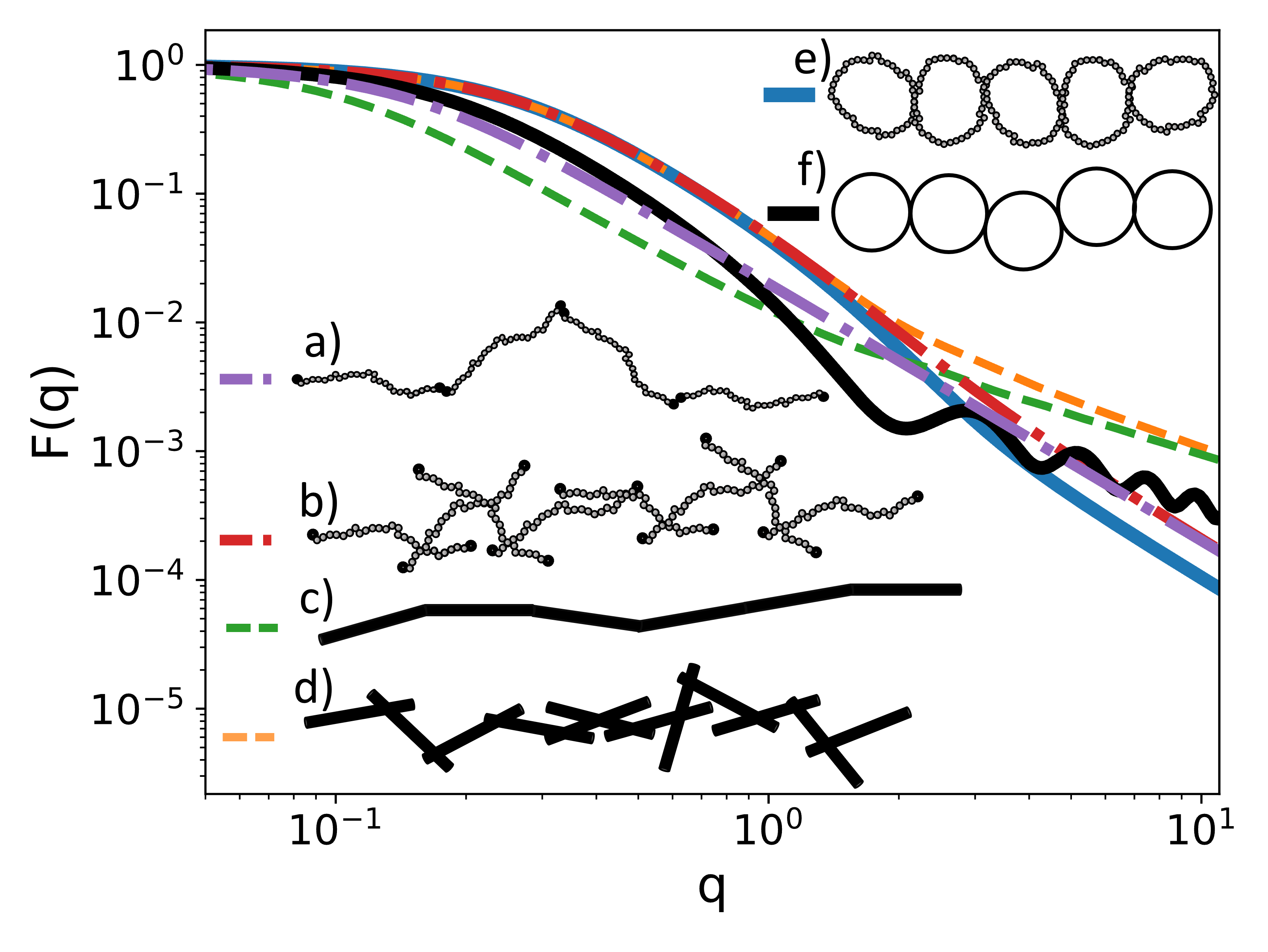}
    \caption{ Scattering from a chain of $N=100$ identical linked sub-units for
     a) "end2" to "end1" linked Gaussian polymers,
     b) "contour" to "contour" linked Gaussian polymers,
     c) "end2" to "end1" linked rods,
     d) "contour" to "contour" linked rods,
     e) "contour" to "contour" linked polymer loops, and
     f) "contour" to "contour" linked circles. The structural parameters of the sub-units are chosen such
     that their radius of gyration is one.
    }
\end{figure}
\twocolumn

\subsection{Chain}

\begin{verbatim}
1: World w;
2: GraphID rw = w.Add("GaussianPolymer","P1","p");
3: for (int i=2; i<=100; i++)
4:    {
5:        string now ="P"+to_string(i)  +".end1";
6:        string last="P"+to_string(i-1)+".end2";       
7:        w.Link("GaussianPolymer",now,last,"p");
8:    }
9:  w.Add(rw, "RandomWalkPolymer");
10: ex F=w.FormFactor("RandomWalkPolymer");
11: ParameterList params;
12: w.setParameter(params,"beta_poly",1);
13: w.setParameter(params,"Rg_poly",1);
14: DoubleVector qvec = w.logspace(0.01, 50.0, 400);
15: w.Evaluate(F, params, qvec, "chain_end2end.q");
\end{verbatim}

To illustrate the versatility of SEB below we show a short C++ program that generates a chain of $100$ identical polymers
linked end-to-end. The program obtains the symbolic expression for the form factor, and then uses several helper methods
to evaluate this equation for specific parameters, finally producing a file with $F(q)$ vs. $q$.

Lines 2-8 creates the chain. Initially we add a single polymer "P1", then we use a for loop to add and link $99$ more polymers.
The polymers have unique names "P(N)", where N denote the number of the sub-unit. The strings variables now and last, is the name of the current and previous sub-units. They are all identical and both are tagged as "poly". The linkage is "P(N).end1" to "P(N-1).end2" for all polymers, such that they form one long continuous chain.  In Line 9, we name this structure "RandomWalkPolymer", and obtain the symbolic expression for its form factor $F$ in Line 10. In Lines 11-13 we define a list of parameters, and set the excess scattering length "beta\_poly" to one, and the radius of gyration "Rg\_poly" is also set to one. In Line 13, we generate qvec, which is a vector of all the $q$ values at which we want to evaluate the form factor. We choose 400 log-equidistant points between $q_{min}=0.01$ and $q_{max}=50$. From the point of view of SEB, units are irrelevant. All scattering expressions depend on dimensionless products of structural length scales and a $q$ value, and as long as both are expressed with a consistent choice of unit, the unit will cancel when evaluating the scattering profile numerically. Finally in Line 15, we evaluate the symbolic expression by inserting the list of parameters and each of the $q$ values in the expression. The result is saved to a file "chain\_end2end.q".

A plot of that file is shown in Fig. \ref{fig:chain100}a. We can now study how the scattering profile changes when we keep the chain structure, but change the sub-unit and/or the linkage. Replacing "GaussianPolymer" by "ThinRod" directly generates a file with the scattering for a chain of rods linked end-to-end. This is shown in Fig. \ref{fig:chain100}c. Replacing "end1" and "end2" by "contour.r(N)" and "contour.s(N-1)" produces the contour-to-contour linkage shown in Fig. \ref{fig:chain100}bdef, where for the latter two curves, we chose "GaussianLoop" or "ThinCircle" as sub-units.

In the Guinier regime of Fig. \ref{fig:chain100}, we observe that the end-to-end linked rods have the largest radius of gyration followed by the end-to-end linked polymers. These form the most loose and extended chain structure. The contour-to-contour linked rods, polymers and loops have the smallest radius of gyration, which is consistent with these chains being the most dense and collapsed structures. Since a chain of $100$ end-to-end linked polymers with $R_g^2=1$ corresponds to a single polymer with $R_g^2=100$ the  scattering is the Debye form factor. At large $q$ values, for all polymer structures we observe the $(qRg)^{-2}$ power law consistent with local random walk statistics. For chains built with rods, we see a $(qL)^{-1}$ power law behavior at large $q$ values which is expected from a rigid rod. The chains of circles structure shows oscillations due to the regular distance between scatterers on a circle, but the trend line of the oscillations follows a $q^{-1}$ power law consistent with local rod-like structure.

\subsection{Dendrimers}

\onecolumn
\begin{figure}\label{fig:dendrimer}
    \includegraphics[width = 0.80\columnwidth]{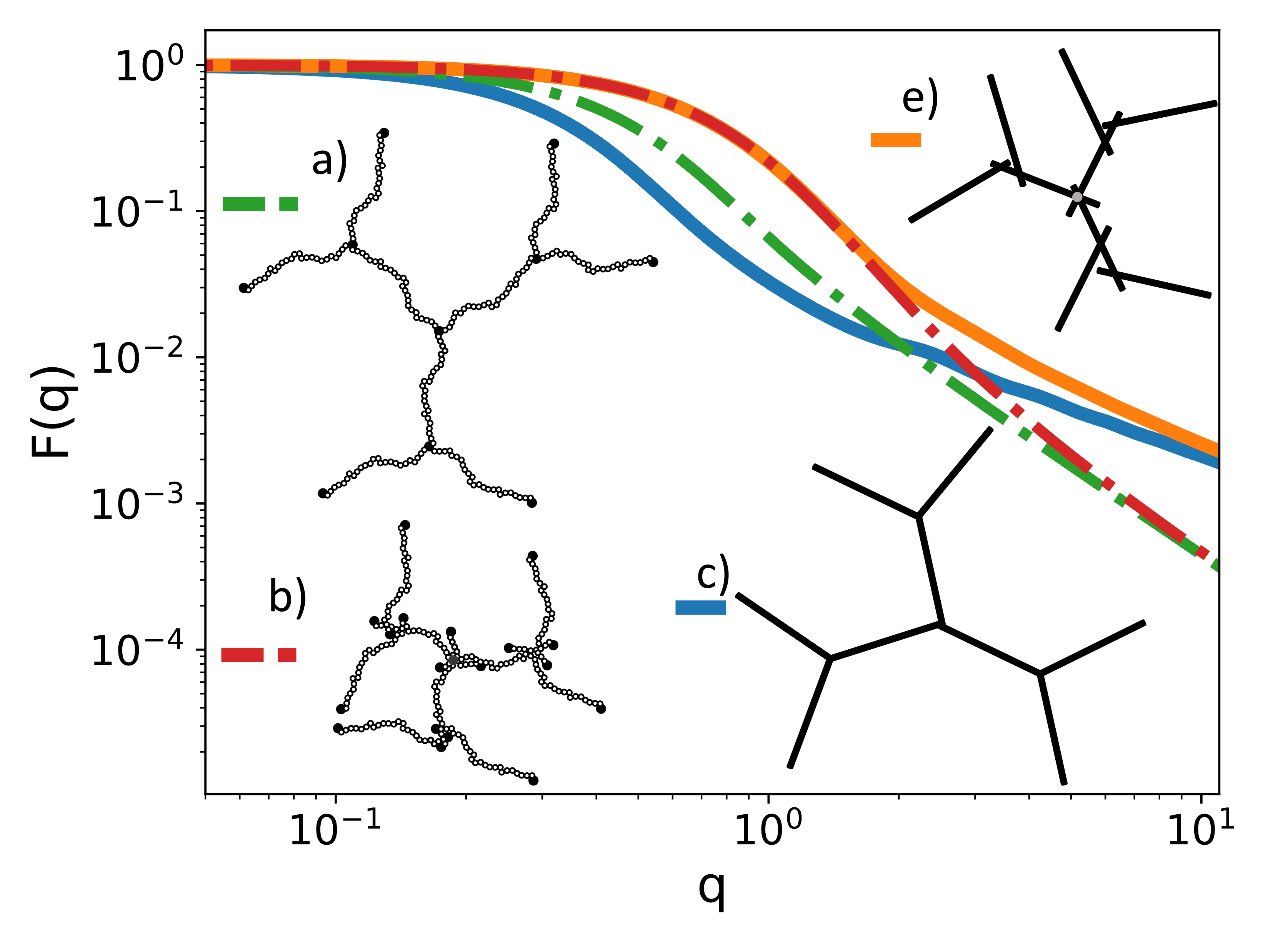}
    \caption{ Scattering from dendrimer with $4$ generations and $3$ functional links.
    a) "end1" to "end2" linked polymers,
    b) "contour" to "contour" linked polymers,
    c) "end1" to "end2" linked rods,
    d) "contour" to "contour" linked rods.
     Structural parameters of sub-units chosen so the radius of gyration is always one. The sketches of the dendrimer structures only show the first two generations for sake of brevity.
    }
\end{figure}
\twocolumn
 
\begin{verbatim}
1: GraphID dendrimer = w.Add("Point","center");
2: int count=0;
3: Attach(4, 3, "center.point", count, w);
4: w.Add(dendrimer, "Dendrimer");
\end{verbatim}

Generating a dendritic structure calls for a recursive function, and the challenge is how to assign names systematically so the links are consistent with a dendritic structure. In line 1 we define a Point, which we call "center". This is an invisible sub-unit with zero excess scattering length, but which is useful as a seed to attach other sub-units to. In line 2, we define a counter which will be counting the number of sub-units added. The recursive function Attach() generates the dendrimer (see code below), and is called in Line 3. The argument $4$ is the number of generations to generate, and $3$ is the functionality of each connection point. The "center.point" is the initial reference point on which to graft additional polymers. The two last arguments are the counter and the world we are adding sub-units into. In the last line we name the resulting structure "Dendrimer". The rest of the code for generating a file with the form factor is identical to the chain example above.

\begin{verbatim}
1:void Attach(int g, int f, string ref, int& c, World& w)
2:{
3: int arm=f-1;
4: if (ref=="center.point") arm=f;
5: for (int i=0;i<arm;i++)
6:  {
7:   string name = "S"+to_string(c)+".end1";
8:   w.Link("GaussianPolymer", name, ref, "poly");
9:   string newref  = "S"+to_string(c)+".end2";
10:  c++;
11:  if (g>1) Attach( g-1, f, newref, c, w);
12: }
13:}
\end{verbatim}

The recursive function receives "g" the number of generations that remains to be attached, "f" the functionality of each link, "ref" which is the reference point from the previous generation onto which we link the current generation. "c" and "w" are a global counter and world, respectively. In line 3-4 we define the numbers of arms to attach to this reference point. Usually this is $f-1$ since we are linking to the tip of an existing branch, however in the special case where we are linking arms to the center.point, we need to add func arms instead. That ensures all connection points have desired functionality.

In lines 5-12 we add the arms and link them to the previous generation. In line 7 we define a name for each new sub-unit "S(c)", and in line 8, we add GaussianPolymer sub-units and link them to the tip of the previous generation. The links are "S(c).end1" to ref, where "ref" is the tip of the last generation of polymers. In line 9, we define the new reference point on which to add the next generation. This reference point is "S(c).end2". Finally in line 10, we increment the counter of sub-units that has been added so far. In case we are not done building, that is if $g$ larger than one, in line 11 we again call the Attach function to attach the next generation to the tip of the current arm, that is to newref. Now with the generation decremented by one, the same functionality.

The resulting structure contains 45 sub-units (3 from the 1st generation, 6 from the 2nd generation, 12 from the 3rd generation, and 24 from the 4th generation) The code above generates the structure plotted in Fig. \ref{fig:dendrimer}a. Again by changing line 8 we can link other sub-units such as thin rods. Changing lines 7 and 9, we can change the reference points from end-to-end to contour-to-contour links. The results are the four curves shown in Fig. \ref{fig:dendrimer}. Again we observe in the Guinier regime, that dendrimers made by end-to-end linked rods and polymers have the largest radii of gyration. We also observe that at large $q$ vectors the power laws $(qL)^{-1}$ for rods, and $(qR_g)^{-2}$ for polymers show what sub-units they are built with. We also observe that contour-to-contour linked structures have the same radius of gyration independently of their sub-unit structure.

\subsection{Solids}

\onecolumn
\begin{figure}\label{fig:solidgeometry}
    \includegraphics[width = 0.80\columnwidth]{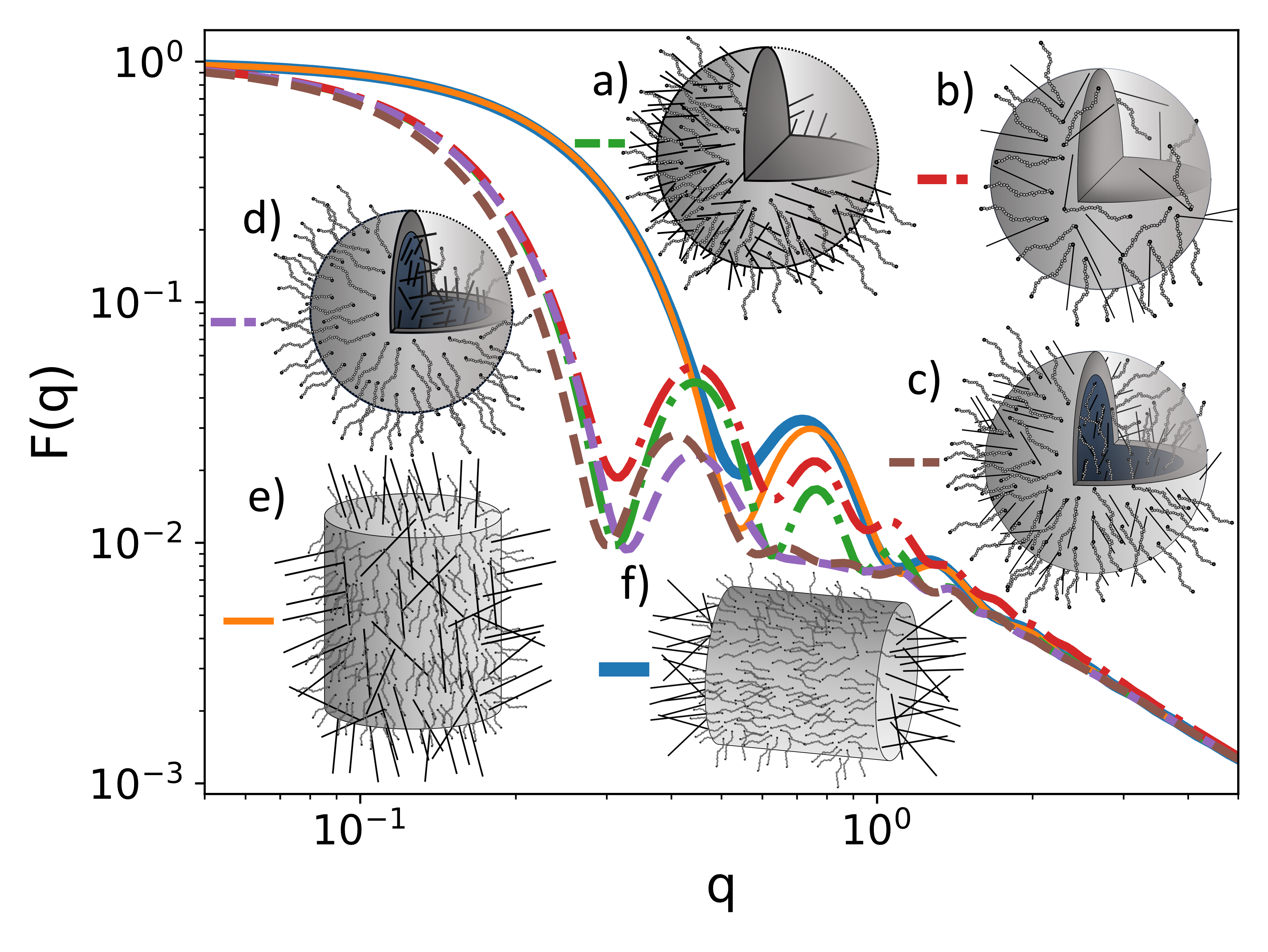}
    \caption{ Scattering from various solid bodies with $50$ rods and $50$ polymers attached
    to different surfaces. The solid body is contrast matched $\beta_{solid}=0$, and $\beta_{poly}=\beta_{rod}=1$.
    a) solid sphere $R=10$ with rods and polymers randomly attached to the surface,
    b) solid sphere $R=10$ with pairs of rods and polymers attached to the same random point,
    c) solid spherical shell $R_i=8$, $R_o=12$ with rods and polymers randomly attached to the interior and exterior surfaces. 
    d) solid spherical shell $R_i=8$, $R_o=12$ with $50$ rods attached to the interior surface and $50$ polymers
    attached to the exterior surface.
    e) cylinder $L=10$, $R=5$ with rods and polymers randomly attached to the surfaces.
    f) cylinder $L=10$, $R=5$ with rods attached to the two cylinder ends and polymers are attached to the hull.  
    For curves c and e where several surfaces contribute area, we have weighted the scattering terms
    with their respective area fractions to ensure homogeneous area coverage in the case of random attachment.
    }
\end{figure}
\twocolumn

\begin{verbatim}
1: GraphID str = w.Add( "SolidSphericalShell", "shell");
2: for (int i=1; i<=50; i++)
3:  {
4:   string name1= "poly"+to_string(i)+".end1";
5:   string ref1 = "shell.surfaceo#p"+to_string(i);
6:   w.Link( "GaussianPolymer", name1, ref1, "poly");
7:   string name2= "rod"+to_string(i)+".end1";
8:   string ref2 = "shell.surfacei#r"+to_string(i);
9:   w.Link( "ThinRod", name2, ref2, "rod");          
10: }
11: w.Add(str, "Structure");
\end{verbatim}

With SEB we can investigate how different linkage options of sub-units on the surface of solid bodies affect the scattering. In the example code above, we generate a solid spherical shell in line 1. The shell is a homogeneous solid body defined by an exterior radius $R_o$ and an interior radius $R_i$. In lines 4-6, we add and link a Gaussian polymer. The polymer is named "poly(i)", and linked by "poly(i).end1" to "shell.surfaceo\#p(i)", where "surfaceo" denotes distributed reference points on the "outer" or exterior surface of the shell. The unique label "p(i)" ensures that all polymers are linked to different random points on the surface. In lines 7-9, we add and link a thin rod. The rod is named "rod(i)", and linked by "rod(i).end1" to "shell.surfacei\#r(i)", where "surfacei" denotes the interior surface. Again the unique label "r(i)" ensures that rods are linked to different random points. In line 11, we name the resulting structure "Structure". As in the chain example, we evaluate the form factor and generate a file with the corresponding scattering curve.

Changing line 1, we can change which solid body we are attaching sub-units to e.g. solid spheres or cylinders. Changing lines 6 and/or 9, we can change what sub-units we link to the surface, and by which reference point the link should be made. Changing the reference points in lines 5 or 8, we can choose different linkage options on the solid body. Fig. \ref{fig:solidgeometry} shows a comparison of some of the possible linkage options. The code above corresponds to the d curve. Here, we choose to contrast match the solid body $\beta_{shell}=0$, and choose $\beta_{poly}=\beta_{rod}=1$. Hence the scattering is due to both the polymers and rods and their  interference contribution which depends on the shape of the body to which they are attached. 

In the Guinier regime of the scattering profiles shown in Fig. \ref{fig:solidgeometry}, we observe that the solid spheres and spherical shells are nearly identical as are the scattering from cylinders. This is not surprising since the scattering between different sub-units is modulated by the phase factor of the solid body on which the sub-units are attached. At very large $q$ values we observe power law behavior with an exponent slightly larger than $-1$. This is to be expected, since the scattering is dominated by the sub-unit form factors, and asymptotically the rod $(qL)^{-1}$ will dominate over the polymer $(qRg)^{-2}$ unless the number of polymers vastly outnumber the number of rods. In the crossover regime, we observe different oscillations for the different linkage options. These oscillations are due to the different distributions of surface-to-surface distances between the tethering points of pairs of rods and / or polymers.

\subsection{Hierarchical structures}

\onecolumn
\begin{figure}\label{fig:chain-of-stars}
    \includegraphics[width = 0.80\columnwidth]{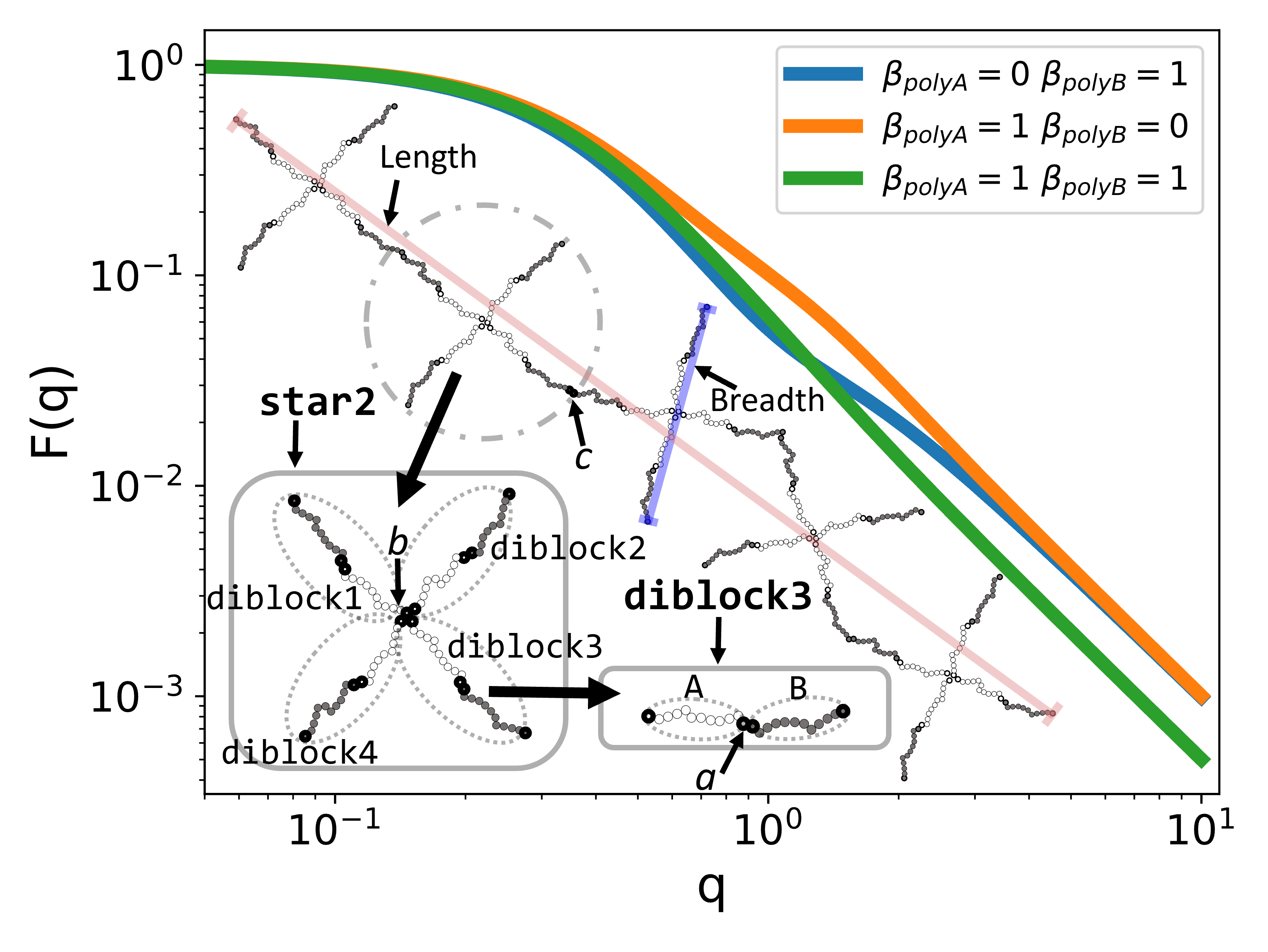}
    \caption{ Scattering from a chain of five four-functional stars where each arm is a diblock copolymer
    for three different choices of contrast. The illustrated links are a) the block copolymer formed by "A.end2" to "B.end1", b) the star formed by "diblock2:A.end1" to "diblock1:A.end1" and similar for the other arms, c) the chain formed by "star2:diblock1:B.end2" to "star1:diblock3:B.end2" and similar for the other stars.
    }
\end{figure}    
\twocolumn

In the examples above we have built structures by connecting sub-units to each other. The result was described by a GraphID, that we could name as a type of structure, and then we could use that name to derive various scattering expressions. Since the formalism is complete any sub-structure can be used as a sub-unit. World has a Link method, that takes a GraphID (referring to a type of structure) and names and links it to an existing structure. This works analogously to Link called with a string denoting a type of sub-unit. The code below illustrates the concept.

\begin{verbatim}
1: GraphID d = w.Add("GaussianPolymer","A");
2: w.Link( "GaussianPolymer","B.end1","A.end2");

3: GraphID s = w.Add(d,"diblock1");
4: w.Link(d,"diblock2:A.end1","diblock1:A.end1");
5: w.Link(d,"diblock3:A.end1","diblock1:A.end1");
6: w.Link(d,"diblock4:A.end1","diblock1:A.end1");

7: GraphID c = w.Add(s,"star1");
8: w.Link(s,"star2:diblock1:B.end2","star1:diblock3:B.end2");
9: w.Link(s,"star3:diblock1:B.end2","star2:diblock3:B.end2");
10:w.Link(s,"star4:diblock1:B.end2","star3:diblock3:B.end2");
11:w.Link(s,"star5:diblock1:B.end2","star4:diblock3:B.end2");

12:w.Add(c, "chain");
\end{verbatim}

In lines 1, we add a Gaussian polymer sub-unit "A", and in line 2 we add and link another Gaussian polymer "B" sub-unit to it as we did several times above. The names "A" and "B" should be thought of as two instantiations of the type of object with an internal structure described by the type "GaussianPolymer". It is important to distinguish between concrete objects of a certain type of structure and the type of structure itself. The type does not exist per se, but is just a generic description. In the case of "A" and "B" these have their own structural parameters, and contribute specific terms to scattering expressions. The type GaussianPolymer is a description of the internal chain statistics of a polymer molecule. When creating a new sub-units or structures in SEB, we instantiate it from a type of structure.  GraphID variables are also types of structure, in particular, the GraphID variable d describes a diblock copolymer structure. In line 3, we add a new structure to the world named "diblock1", which is an instantiation of the diblock type. Hence "diblock1" is a concrete structure in the same sense as "A" and "B" are concrete sub-units.

In lines 4-6, we do something new, we call Link(), not with a sub-unit type, but with the diblock type (GraphID variable d). We name these four new structures "diblock2", "diblock3", "diblock4", respectively. Each structure is linked by a reference point inside the structure to a reference point that already exists in the world. For the diblock2 structure, we link "diblock2:polyA.end1" to "diblock1:polyA.end1", since "diblock1" already exists in the world we can link to it. To link structures, we need to specify the path to get from the structure level via sub-structures down to the reference point, which is associated with a specific sub-unit. Since all names are unique, so is any path from a sub-structure to a reference point. The resulting structure is a 4-armed diblock copolymer star, where the "A" blocks are linked by their "end1" reference points and forms the center of the star, while the corona is formed by the four "B" blocks and their free chain ends are at the "end2" reference points. 

While we usually define the GraphID by the return value of the first Add() method, all subsequent Link() calls also return the same GraphID value, since this is associated with the whole graph first created by Add(), and then grown each time the Link() is called. In line 3, we stored the type of graph formed by "diblock1" to "diblock4" in the GraphID variable s, which is now the type of a 4-functional diblock star structure.

In line 7, we now instantiate a star sub-structure and name it "star1". This defines a new GraphID, we save in a variable c. Then in lines 8-11 we proceed to instantiate 4 more star sub-structures named "star2" to "star5". Each time we link "star(n):diblock1:B.end2" to "star(n-1):diblock3:B.end2", since "star(n-1)" already exists, and has a "diblock3:B.end2" reference point inside it. The result is a linear chain of stars formed by linking the tips of "diblock1" and "diblock3", hence "diblock2" and "diblock4" form dangling ends analogously to a bottle-brush structure. Finally to calculate the form factor of this type of chain, we must name it to instantiate it in the world. The rest of the code is similar to the chain example above.

This example illustrates the power of building structures using more simple sub-structures as building blocks. With $15$ lines of code, we have generated a hierarchical structure with $40$ sub-units. Fig. \ref{fig:chain-of-stars} shows an illustration of the resulting structure together with the form factor evaluated for three different contrast options. In the Guinier regime, we observe that the radius of gyration is nearly the same independently of contrast which we would also expect for such a structure. At large $q$ values we obtain the characteristic power law of polymer sub-units. For intermediate $q$ values, the structure is slightly different. When the "polyA" blocks are contrast matched $\beta_A=0$, they play the role of invisible spacers inside the stars. When the "polyB" blocks are contrast matched, they play the role of invisible spacers between different stars.

Besides calculating scattering expressions SEB can also provide expressions characterizing the size of a structure. For instance, w.RadiusOfGyration2("chain") returns an expression for the radius of gyration by applying a Guinier expansion of all sub-unit scattering terms. After simplification, the result is

\begin{equation*}
\langle R_g^2\rangle = 
(\beta_A+\beta_B)^{-2} \left( 12.9  R_{g_{A}}^{2} \beta_{B}^{2}+9.6  R_{g_{B}}^{2} \beta_{A}^{2}  \right.
\end{equation*}
\begin{equation*}
\left. +21  R_{g_{B}}^{2} \beta_{A} \beta_{B}+24.3  R_{g_{A}}^{2} \beta_{A} \beta_{B}+11.3  R_{g_{B}}^{2} \beta_{B}^{2}+11.3  R_{g_{A}}^{2} \beta_{A}^{2}   \right)
,
\end{equation*}
while the radius of gyration measures the distances between all pairs of scatterers, we could for instance also ask what is the mean-square distance between the center of the star and all scatterers in the structure. A Guinier expansion of the corresponding form factor amplitude provides the result, and "star3:diblock1:polyA.end1" is the reference point at the center of the star, hence this mean-square distance gives an idea of the radial extent of the structure. Calling w.SMSD\_ref2scat( "chain:star3:diblock1:polyA.end1") returns that result. The method is called SMSD for $sigma$ mean-square-distance to remind the user to account for a potential symmetry factor. Finally, we could ask what is the length and breadth of the structure. To calculate the length, we call w.SMSD\_ref2ref( "chain:star1:diblock1:polyB.end2", "chain:star5:diblock3:polyB.end2") which returns the mean-square distance between
the two reference points at either end of the structure. The result is
$\langle R^2_{length}\rangle=60 ( R_{g_{B}}^{2}+R_{g_{A}}^{2})$
 To estimate the breadth of the structure, we change the reference points to 
w.SMSD\_ref2ref( "chain:star3:diblock2:polyB.end2", "chain:star3:diblock4:polyB.end2"), since "diblock2" and "diblock4" are the two dangling diblocks, and the "polyB.end2" are the dangling ends of these diblocks. The result is
$\langle R^2_{breadth} \rangle=12 ( R_{g_{B}}^{2}+R_{g_{A}}^{2})$. 
 
These results are easy to obtain by hand. Noting for a single polymer $R_g^2(N) = \langle R^2_{end2end}\rangle / 6 = b^2 N /6$  where $b$ is the random walk step length, and $N$ number of steps in the polymer. Then to estimate the number of steps along the length of the chain, we note that it has $10$ A blocks and $10$ B blocks from one end to the other. Hence $\langle R^2_{length}\rangle $$= b^2 N_{length} $$= b^2(10 N_A+ 10 N_B) $$= 60 (R_{g_{A}}^2 +R_{g_{B}}^2)$. For the breadth, note a star has a breadth of $N_{breadth}=2N_A+2N_B$. The result is that the chain is five times longer than its breadth, which is what one would expect.

\section{Summary}\label{sec:summary}

The main problem in analyzing small-angle scattering (SAS) data is the availability of model expressions for fitting. 
Here we presented the "Scattering Equation Builder" (SEB) which is an open-source C++ library available
at Ref. \cite{SEBgithub}. SEB automates part of this problem by generating symbolic expressions for complex
composite models of structures using the formalism presented in Refs. \cite{scatterformalism,scatterformalism2}.
The formalism is built on the assumption that sub-units are mutually non-interacting, and the assumption that structures
do not contain loops. Finally all links are assumed to be completely flexible. No further mathematical simplifications or
approximations are made. In particular, no assumptions are made regarding the internal structure of sub-units.

With SEB users write short programs that construct a structure using sub units and simpler structures as building blocks. 
Much like LEGO, sub-units can be linked at certain points called reference points. These can be either specific
geometric points such as one of the ends of a polymer, or they can be randomly distributed e.g. on the surface of a sphere.
With the building blocks of sub-units and reference points, a large number of complex structures can be built with relative ease. See Fig. \ref{fig:subunitoverview} for the sub-units and reference points supported by this initial release.

SEB derives analytic symbolic expressions for the form factor, form factor amplitude, phase factor of a structure.
SEB can also derive expressions for the radius of gyration as well as the mean-square distance between a reference
point and all scatterers in a structure. Finally SEB can derive the mean-square distance between pairs of reference points.
The expressions can evaluated to a number e.g. when fitting, evaluated to produce a file for plotting, or they can be
outputted several formats for LaTeX documentation, C/C++ and python compatible equations, or exported to matlab / Mathematica. 

In the present article, we have given simple illustrative examples as well as some more complex examples of what SEB can do.
SEB is available at GitHub \cite{SEBgithub}, and a frozen version related to the present work is deposited on Zenodo\cite{SEBZenodo}, 
We hope the SEB library will grow as more sub-units becomes supported, and we welcome contributions from the users in developing future versions of the library.

%\ack{Acknowledgements}

     % References are at the end of the document, between \begin{references}
     % and \end{references} tags. Each reference is in a \reference entry.

%\begin{references}
%\reference{Author, A. \& Author, B. (1984). \emph{Journal} \textbf{Vol},  first page--last page.}
%\end{references}

\end{document}